\documentclass[journal]{IEEEtran}
\ifCLASSINFOpdf
\else
\fi
\usepackage[dvipsnames]{xcolor}
\usepackage{tikz}
\usetikzlibrary{shapes.geometric, arrows}
\tikzstyle{startstop} = [rectangle, rounded corners, minimum width = 3cm, minimum height =1cm, text centered, draw = black, fill = NavyBlue!50]
\tikzstyle{io} = [ trapezium, trapezium left angle=70, trapezium right angle = 110, minimum width=3cm, minimum height =1cm, text centered, draw = black, fill = blue!30]
\tikzstyle{process}= [rectangle, minimum width=3cm, minimum height = 1cm, text centered, draw = black, fill= blue!30]
\tikzstyle{decision} = [diamond, minimum width = 3cm, minimum height=1cm, text centered, draw=black, fill = green!30]
\tikzstyle{arrow} = [thick,->,>=stealth]

\usepackage{circuitikz}
\usepackage{amsmath}
\usepackage{subcaption}
\usepackage{lipsum}
\usepackage{graphicx}

\usepackage{hyperref}

\begin{document}
%
\title{Analysis and Design of a PMUT-based transducer for Powering Brain Implants}
%
%
%

\author{Fernanda~Narv\'aez,~\IEEEmembership{Student Member,~IEEE,}
        Seyedsina~Hosseini,~\IEEEmembership{Graduate Student Member,~IEEE,}
       Hooman~Farkhani,~\IEEEmembership{ Member,~IEEE,}
       and~Farshad~Moradi,~\IEEEmembership{Senior Member,~IEEE}
\thanks{The authors are with the Integrated Circuits and Electronics Laboratory (ICE-Lab) Department of Engineering, Aarhus University, 8200 Aarhus N, Aarhus, Denmark (e-mail:{\tt\small ; Moradi@eng.au.dk}).}
\thanks{Manuscript received ; revised ...}}

%
%

\markboth{Journal of }%
{Shell \MakeLowercase{\textit{et al.}}: Bare Demo of IEEEtran.cls for IEEE Journals}
%



\maketitle

\begin{abstract}
\textbf{This paper presents an analytical design of an ultrasonic power transfer system based on piezoelectric micromachined ultrasonic transducer (PMUT) for fully wireless brain implants in mice. The key steps like the material selection of each layer and the top electrode radius to maximize the coupling factor are well-detailed. This approach results in the design of a single cell with a high effective coupling coefficient. Furthermore, compact models are used to make the design process less time-consuming for designers. These models are based on the equivalent circuit theory for the PMUT. A cell of 107 $\mu$m in radius, 5 $\mu$m in thickness of Lead Zirconate Titanium (PZT), and 10 $\mu$m in thickness of silicon (Si) is found to have a 4$\%$ of effective coupling coefficient among the highest values for a clamped edge boundary conditions. Simulation results show a frequency of 2.84 MHz as resonance. In case of an array, mutual impedance and numerical modeling are used to estimate the distance between the adjacent cells. In addition, the area of the proposed transducer and the number of cells are computed with the Rayleigh distance and neglecting the cross-talk among cells, respectively. The designed transducer consists of 7x7 cells in an area of 3.24 mm$^2$. The transducer is able to deliver an acoustic intensity of 7.185 mW/mm$^2$ for a voltage of 19.5 V for powering brain implants seated in the motor cortex and striatum of the mice's brain. The maximum acoustic intensity occurs at a distance of 2.5 mm in the near field which was estimated with the Rayleigh length equation.}
\end{abstract}


%
\IEEEpeerreviewmaketitle

\section{INTRODUCTION}
%
%
%
%
\IEEEPARstart{T}{he} evolution of technology has allowed that implantable medical devices reach astounding miniaturization. Implants interfacing the human brain aid in the diagnosis, monitoring, and treatment of chronic brain disorders such as Parkinson’s Disease (PD), depression, and Alzheimer's \cite{famm2013jump}. Nowadays, Optogenetics emerges as a treatment for PD. This method delivers precisely lights to specific brain regions and stimulates neurons circuits \cite{deisseroth2015optogenetics}. In this research field, a fully implantable battery-less and wireless controlled device is a promising alternative \cite{rashidi2019stardust}. However, methods for powering such implants remain in continuous research. The main development in wireless powering for brain implants has primarily been based on coupled coils \cite{ho2014wireless,xu2018optimization,mita2018microscale}. Despite its wide use, electromagnetic waves have poor propagation through tissue due to attenuation. In addition, there is a high risk of tissue overheating and interference from surrounding devices \cite{li2018feasibility}. To solve this issue, brain implants have recently included a piezoelectric receiver to harvest energy, which utilizes an ultrasound wireless transfer system to have a continuous power source \cite{Dagdeviren1927}\cite{xu2010piezoelectric}.

In general, commercial ultrasound transducers are built with bulk piezo-ceramic materials. Miniaturization of bulk transducers has not evolved enough due to the difficulty in the manufacturing process \cite{khuri2009next}, although recently top-electrode-patterned bulk transducers have been realized for powering brain implants in different locations in the brain \cite{hosseini2020s}, \cite{hosseini2019multi}. Furthermore, the implementation with integrated circuit becomes a challenge that leads to unmet needs in the field of powering for freely moving animals. On the other hand, MEMS-based transducers, i.e. piezoelectric micromachined ultrasonic transducer (PMUT) and capacitive micromachined ultrasonic transducer (CMUT) have shown promising results in ultrasound imaging \cite{smyth2017piezoelectric} \cite{wygant2008integration}, and in recent years in powering applications. In addition, they have low acoustic impedance and cost-effective manufacturability \cite{khuri2009next}. In the case of CMUT-based transducer, the power consumption increases due to the required bias voltage, giving an advantage to PMUT over CMUT \cite{jung2017review}. 

\begin{figure}[h!]
    \centering
    \includegraphics[width=0.5\textwidth]{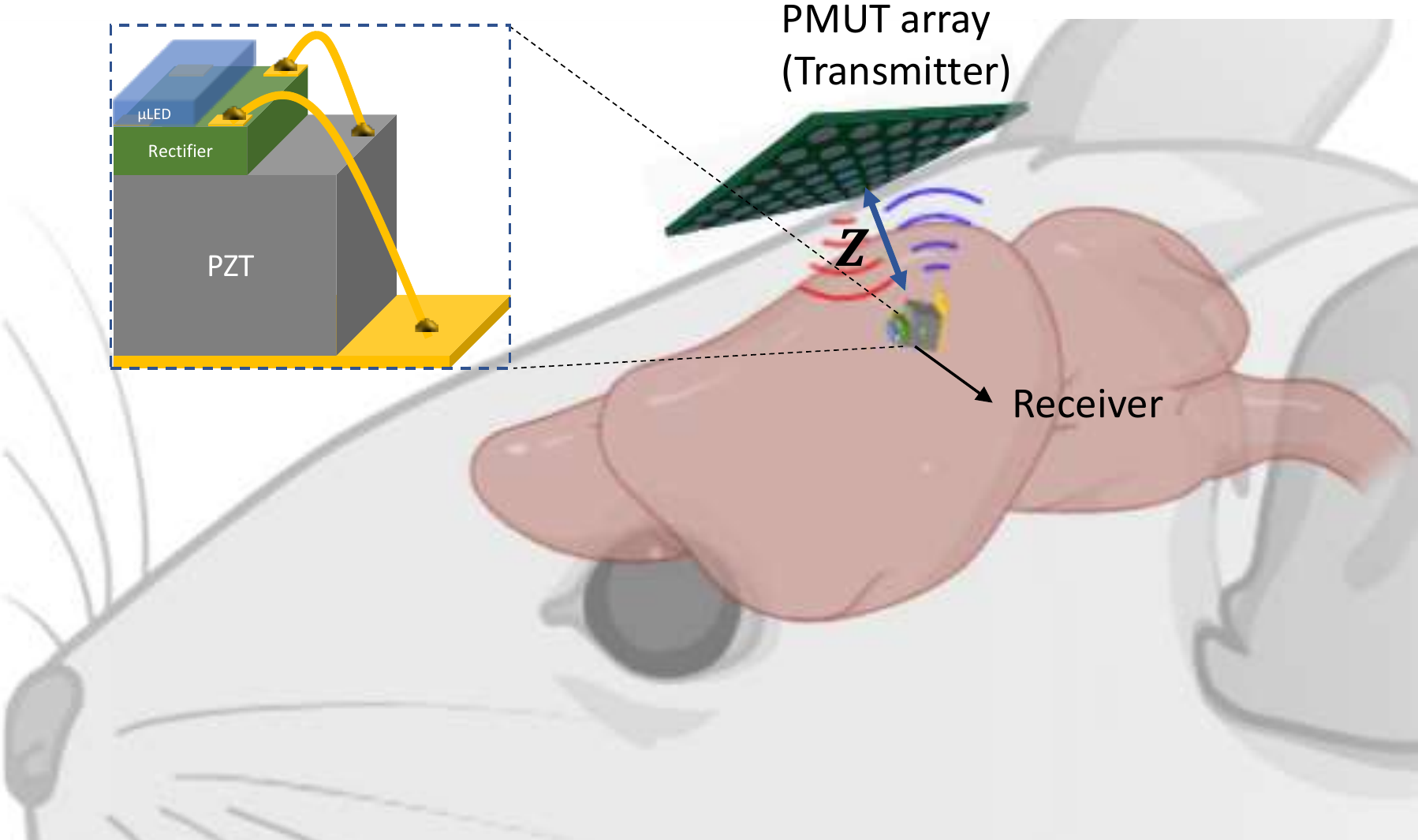}
    \caption{Schematic for power transfer to a brain implant mounted at $Z$ distance between 1 to 4mm where the mice's motor cortex and striatum are located.}
    \label{fig:scheme_wirelessPT}
\end{figure}

This article presents the design of a PMUT-based transducer by using compact equations that improve the predictability of the output performance of the array. By focusing on specific requirements such as the location, the maximum acoustic intensity, and operation frequency of the receiver, this work brings a PMUT array technology for powering implantable circuitry. The conceptual scheme for the wireless power transfer system is illustrated in Fig. \ref{fig:scheme_wirelessPT}, where a PMUT-based array is proposed as an ultrasonic source (transmitter). The receiver is a brain implant applicable for Optogenetics \cite{stirman2011real} or neuromodulation in rodents \cite{tufail2010transcranial} which  is recently developed by the authors \cite{laursen2020ultrasonically} stacked up with $500\mu m \times 500 \mu m \times 500\mu m$ Lead Zirconate Titanium (PZT) cube as an ultrasonic energy harvester, $300 \mu m \times300 \mu m$ electronic chip as voltage rectifier, and $300\mu m \times130\mu m$ Light Emitting Diode ($\mu$LED) for doing Optogenetics located in a distance between 1 to 4 mm labeled as $Z$ where the motor cortex and striatum of the mice are located. The receiver has a resonance frequency of 2.7 MHz because of the PZT's dimension. The maximum value of the acoustic intensity is limited by FDA-regulations at 7.2 $mW/mm^2$ \cite{fda2008guidance}.

\
Starting with an insight into the equivalent circuit model of the PMUT-based transmitter in section \ref{sec:Equivalen_circuit}, the functional relationships between lumped elements and design parameters are presented. This section includes the formulation of the effects in the resonance frequency of the PMUT in the presence of the brain tissue. The equivalent circuit model allows us to arrive at the design process of a single cell PMUT in section \ref{sec:Design_process} where the effective coupling coefficient is used as a figure of merit for a force response. For the resonance frequency, the diaphragm radius is determined and the cell design is completed. This cell is used to develop the wireless power transfer system design in \ref{subsec:WPT}. The design is verified by simulations in COMSOL V5.4 as it is described in section \ref{sec:Simutation_res}.

\section{EQUIVALENT CIRCUIT MODEL \label{sec:Equivalen_circuit}}
This section deals with the analytical representation of the single and multi-cell configurations of a PMUT that can be used for any application. Starting with an introduction to the derivation of equations for deflection and electrode coverage is presented in addition to the analytical equivalent circuit based in \cite{smyth2017piezoelectric}. The latter provides a visual representation of individual parts and interconnections of the transducer allowing a direct calculation of its geometrical parameters. 
\subsection{Single Cell Model}
\begin{center}
\begin{figure}[h!]
\centering
  \begin{subfigure}[t]{0.4\textwidth}
  \centering 
        \includegraphics[width=\textwidth]{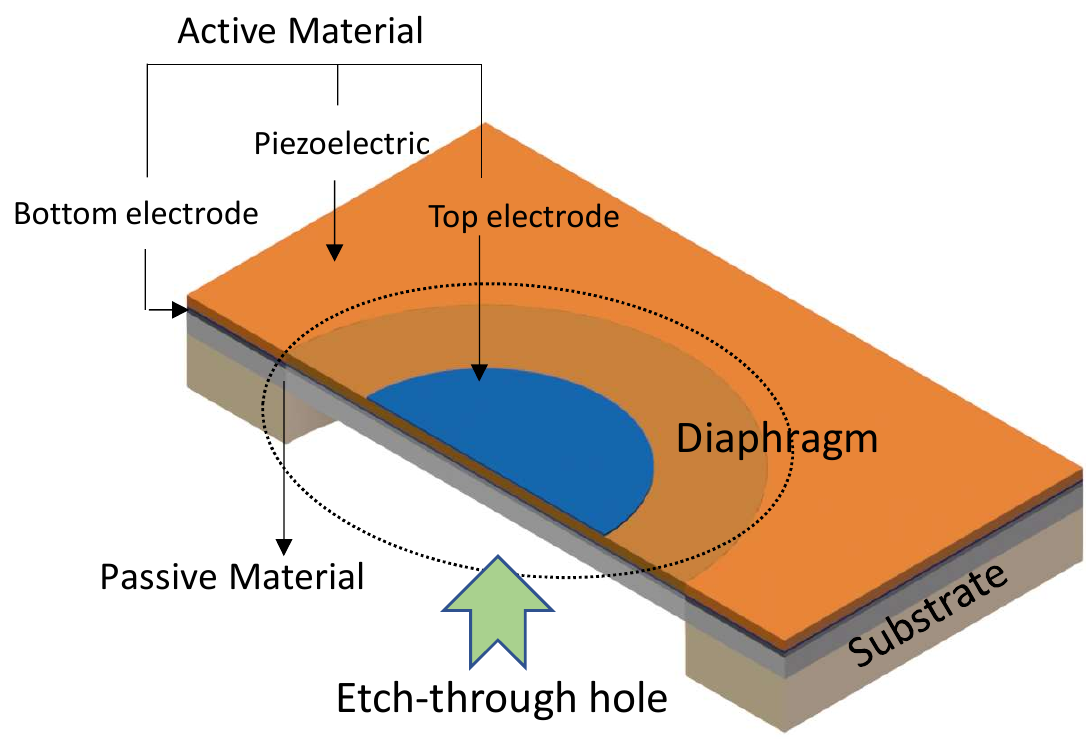}
        \caption{ } \label{fig:schematic_pmut}
    \end{subfigure}
    \begin{subfigure}[t]{0.5\textwidth}
        \centering 
        \includegraphics[width=\textwidth]{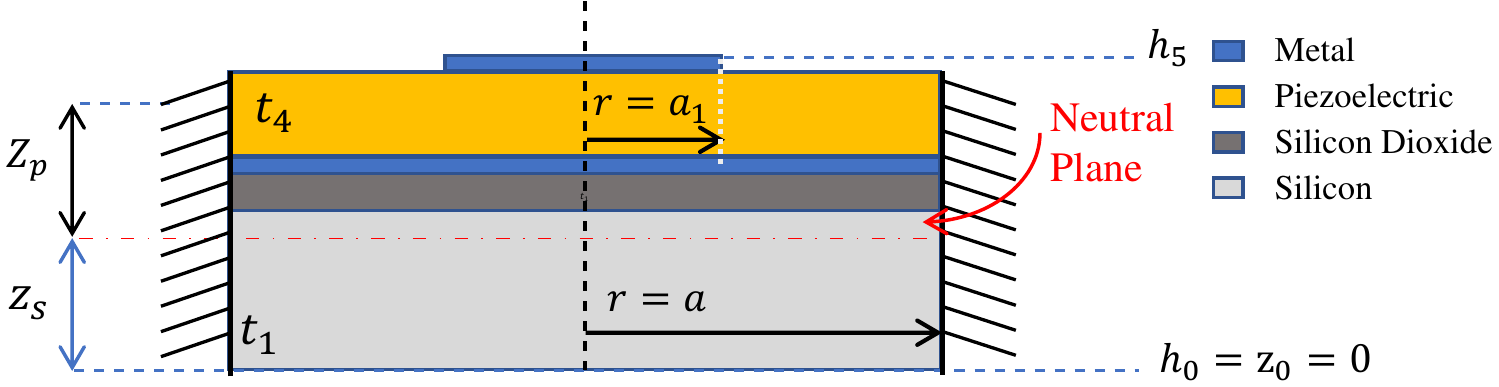}
        \caption{}\label{fig:cross_view}
    \end{subfigure}
    \caption{Schematic of a unimorph PMUT. (a) Single cell view of a PMUT. (b) Cross section of PMUT's diaphragm.}
    \label{fig: PMUT_schematic}
  \end{figure}
\end{center}

Fig. \ref{fig: PMUT_schematic} presents a schematic of the PMUT structure which consists of a single layer or multi-layer \cite{akhbari2015bimorph} of active piezoelectric material sandwiched between two thin metal layers, and a passive part which consists of silicon dioxide ($SiO_2$) and silicon layers.\

\begin{figure}[h!]
\begin{center}
\begin{circuitikz}[scale=0.7][american voltages]
\draw
  (0,0) ++ (0,4) coordinate(Pos)
  to [short, *- ,i=$I$] ++(2,0) coordinate(Cap)
  to [C, l=$C_{01}$] ++(0,-4)
  to [short,-*] (0,0) 
  (Cap) to ++(2,0) coordinate(Pri)
  to [L] ++(0,-4)
  to ++(-2,0)
  (Pos) to [open, v^>=$V_{in}$] (0,0)
  (Pri) ++(1,-4) coordinate(Sec)
  to [L] ++(0,4)
  to ++(1,0)
  to [C, l_=$C_{m,1}$] ++(2,0)
  to [L, l_=$L_{m,1}$] ++(2,0)
  to [short, -* ,i=$u$] ++(1,0)
  to [L, l_=$L_{r,1}$] ++(0,-2)
  to [R, l_=$R_{r,1}$] ++(0,-2)
  to [short,*-] (Sec)
  (Pri) ++(-0.25,0.5)
  node[anchor=west] {$1:N_1B_{11}$};
  \end{circuitikz}
  \end{center}
\caption{Mason's equivalent circuit model of the piezoelectric transducer.}
\label{fig:EquivalentCircuit}
\end{figure}
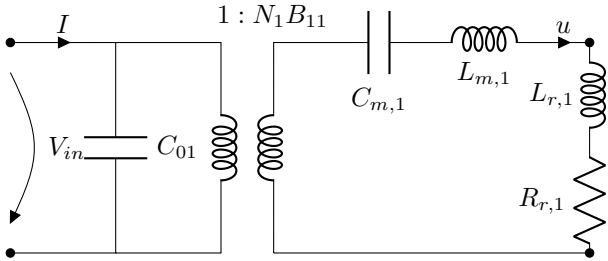

When an alternating voltage is applied out-of-plane, in-plane stresses are generated across the piezoelectric layer, bringing about perpendicular displacements \cite{jung2017review},\cite{dangi2017system}. Classical plate formulation for a single layer circular plate can be assumed if the top and bottom electrodes are uniformly spread, and deposited very thin compared to the overall structure. Following the derivation of \cite{smyth2013analytic}, the governing plate vibration is described in (\ref{eq:governing_plate_vibration}) that results from the integration of plate bending moments:\\
\begin{equation}
D \nabla^2 \nabla ^2 w + \mu \frac{\partial^2 w}{\partial t^2} = \nabla^2 M_p + P_{ext}
\label{eq:governing_plate_vibration}
\end{equation}

\noindent where $w$ is the deflection in the $z$ direction, $M_p$ is the piezoelectric moment, $P_{ext}$ is an externally applied, transverse pressure. $D$ is the flexural rigidity derived by \cite{muralt2005piezoelectric} and defined in (\ref{eq:FlexuralRigidity}) for a five-layer structure.

\begin{equation} 
   D = \frac{1}{3} \sum_{n=1}^5 \frac{(h_n - z_s)^3 - ( h_{n-1} - z_s)^3}{s_{11}^{(n)}(1 - (\nu_{12}^{(n)})^2)}
   \label{eq:FlexuralRigidity}
\end{equation}\

\noindent where $z_s$ represents the neutral axis, $s_{11}^{(n)}$ is the compliance at constant electric field and $\nu_{12}^{(n)}$ is the Poisson ratio of the $n$th layer. The surface density $\mu$ is computed in (\ref{eq:SurfaceDensity}) as

\begin{equation}
    \mu = \sum_{n=1}^5 \rho_n t_n
    \label{eq:SurfaceDensity}
\end{equation}\

\noindent where $\rho$ is the density and $t$ the thickness. The solution to the differential equation is defined in \cite{smyth2017piezoelectric} by using the Green's function and the right-hand side of (\ref{eq:governing_plate_vibration}) as $f(r)=\frac{1}{D}(\nabla^2 M_p + P_{ext})$. In (\ref{eq:BendingMoment}), the piezoelectric bending moment $M_p$ is attained for a single electrode structure:

\begin{equation}
    M_p = e_{31,f} Z_p V_N(t) [H(r)-H(r-a_1)]
    \label{eq:BendingMoment}
\end{equation}\

\noindent where $Z_p$ is shown in Fig. \ref{fig: PMUT_schematic}(\subref{fig:cross_view}) and defines the distance between the neutral axis and the middle of the piezoelectric layer. $e_{31,f}$ is the modified transverse piezoelectric constant \cite{jung2017review}, $V_N(t)$ is the applied sinusoidal voltage and H is the Heaviside step function. For this analysis only axisymmetric modes and the first vibration mode to maximize the deflection are taken into account. The complete solution for ($\ref{eq:governing_plate_vibration}$) is defined in (\ref{eq:BendingMomentfreq}) in the frequency domain as:

\begin{align}
    W(r_0) =& \frac{e_{31,f} Z_p}{D} \frac{V}{\Lambda_{01}(\gamma_{01}^4 - \gamma^4)} \Psi_{01}(r_0) B_{11} \nonumber\\[0.5em]
    +& \frac{P_\omega}{D} \frac{1}{\Lambda_{01}(\gamma_{01}^4 - \gamma^4)} \left [ \frac{2aJ_1(\lambda_{01})}{\lambda_{01}}\right ] \Psi_{01} (r_0) 
    \label{eq:BendingMomentfreq}
\end{align}\

\noindent where $r_0$ is considered as the point where the force is applied, $\Psi_{01}$ is the characteristic shape profile. $P_\omega$ is the frequency-dependent pressure. $a$ is the radius of the PMUT diaphragm as it is illustrated in Fig. \ref{fig: PMUT_schematic}(\subref{fig:cross_view}). $\gamma_{01}$ is the normalized mode shape variable, $\Lambda_{01}$ is a deflection profile constant, the frequency-dependent term $\gamma^4 = \dfrac{\omega^2 \mu}{D}$, and $B_{11}$ is the first parameter of the equivalent circuit shown in Fig. \ref{fig:EquivalentCircuit}, which represents a dimensionless parameter for the coupling between the active area of the electrode and the deflection \cite{sherman2007transducers}. In (\ref{eq:CouplingFactor}), the coupling factor for a single electrode is defined:  

\begin{equation}
    B_{11} =\gamma_{01} \left [ a_{ac} \left( J_1(\gamma_{01} a_{ac}) + \frac{J_0(\lambda_{01})}{I_0(\lambda_{01})}  I_1(\gamma_{01}a_{ac}) \right) \right ] 
    \label{eq:CouplingFactor}
\end{equation}\

\noindent where $J$ and $I$ are the Bessel's function. $a_1$ is the radius of the top electrode as it is depicted in Fig.\ref{fig: PMUT_schematic}(\subref{fig:cross_view}). Lastly, $a_{ac} = \dfrac{a_1}{a}$ is the ratio between radius of the top electrode and the membrane. 

By taking the approach of \cite{smyth2017piezoelectric} and Mason's equivalent circuit, three domains are identified: electrical, mechanical, and acoustic. The first corresponds to the piezoelectric, and it is given by the clamped capacitor $C_{01}$. The coupling between the electrical and mechano-acoustic domains is given by an ideal transformer $1:B_{11}N_1$. Then, the mechanical domain is given by the inductive element $L_{m,1}$ which represents the mass, and the $C_{m,1}$ capacitor that represents the stiffness. The equivalent circuit parameters are defined as:
\begin{align}
    C_{01} & =  \frac{\epsilon_{33}^\sigma a_1^2}{t_p} (1-k_{31}^2)\\
    N_1 & = 19.526 e_{31,f} Z_p\\
    C_{m,1} & = \frac{a^2}{197.526 \pi D}\\
    L_{m,1} & = b_1 \pi a^2 \mu 
\end{align}
\noindent where $\epsilon_{33}$ is the dielectric constant. $b_1$ is defined in \cite{smyth2017piezoelectric} as the boundary condition mode shape constant given by (\ref{eq:constant_boundarycondition}).
\begin{equation}
    b_1 = \left[ \frac{1}{2} \left ( \frac{\lambda_{01}J_0(\lambda_{01})}{2J_1(\lambda_{01})}\right )^2 \right ] 
    \label{eq:constant_boundarycondition}
\end{equation}

The last lumped element is the acoustic impedance, and it is given for two cases for $ka>>1$ where $k$ is the wavenumber and the acoustic impedance is purely resistive, and obtained by the characteristic impedance times the area of the transducer. The second case is when $ka<<1$ and the radiation impedance is defined in (\ref{eq: radiation_impedance}) as:

\begin{equation}
    Z_{r,01} = \pi a^2 \rho_0 c  \left( \frac{(ka)^2}{2} + i ka \right) 
    \label{eq: radiation_impedance}
\end{equation}\

\subsection{Natural response and fluid effects}
The resonance frequency for PMUT depends on the shape, dimensions, and mechanical stiffness of the diaphragm, unlike bulk piezoelectric transducers whose resonance frequency is defined by the thickness. In addition, it is crucial for medical applications to consider the effect of the acoustic impedance of the human body in the resonance frequency. This is usually modeled as a plate, loaded on one side by a fluid. When a structure is in contact with a fluid, the expression to calculate the resonance frequency is corrected in \cite{smyth2017piezoelectric}, \cite{kwak1991axisymmetric}, \cite{woltjer2016optimization} through the mathematical expression in (\ref{eq:frequency}) by adding a virtual mass represented by $\beta$:

\begin{equation}
    f_{f}= \frac{\lambda_{01}^2}{2 \pi a^2} \sqrt{\frac{D}{\mu}} \frac{1}{\sqrt{\beta + 1}}
    \label{eq:frequency}
\end{equation}\

\noindent where $\lambda_{01}$ is the mode shape constant for a defined boundary condition and $\beta=\Gamma \dfrac{\rho_f}{\rho_p}\dfrac{a}{t}$ represents the added virtual mass incremental factor in fluid side of the transducer \cite{kwak1991axisymmetric}. $\Gamma$ is the non-dimensional added virtual mass incremental corrector factor (NAVMI) that depends on $ka$. $\frac{\rho_f}{\rho_p}$ is the ratio between the density of the fluid $\rho_f$ and the plate $\rho_p$. For small values this can be constant as it is derived in \cite{smyth2017piezoelectric}, \cite{kwak1991axisymmetric}, and \cite{olfatnia2011medium}. On the other hand, it is important to define the density of the plate since a PMUT is a multi-layer structure. This paper follows the approach in \cite{sammoura2013optimizing} given by the expression of (\ref{eq:Beta}): 

\begin{equation}
    \beta = \Gamma \frac{\rho_f}{\rho_{Si}} \cdot \frac{a}{t_{Si} + T_{corr,\rho}}
    \label{eq:Beta}
\end{equation}\

\noindent where $T_{corr,\rho}$ is given by (\ref{eq:WithoutSilicon}) and represents the total thickness of all the layers on top of the silicon corrected for its mass density:

\begin{equation}
    T_{corr,\rho} = \sum_{i=2}^5 \frac{\rho_i}{\rho_{Si}} t_i
    \label{eq:WithoutSilicon}
\end{equation}\

\subsection{Multi-cell Model}
The model is based on the idea of a parallel connection of cells. Each cell has its equivalent circuit with lumped elements calculate individually. This is illustrated by the equivalent circuit depicted in Fig. \ref{fig:EquivalentCircuit_array} and the mathematical expression for the acoustic impedance is detailed in (\ref{eq: acoustic_impadance_array}) \cite{oguz2013equivalent}, \cite{akhbari2016curved}:
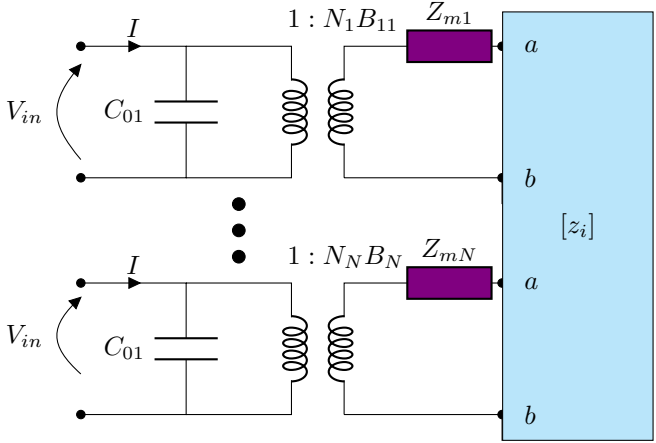
\begin{figure}[h!]
\begin{center}
\begin{circuitikz}[scale=0.7][american voltages]
\draw
  (0,0) to [open, v^>=$V_{in}$] (0,2.5) 
  to [short, *- ,i=$I$] ++(2,0) coordinate(Cap)
  to [C, l_=$C_{01}$] ++(0,-2.5)
  to [short,-*] (0,0) 
  (Cap) to ++(2,0) coordinate(Pri)
  to [L] ++(0,-2.5)
  to ++(-2,0);
 \draw
 (Pri) ++(1,-2.5) coordinate(Sec)
  to [L] ++(0,2.5)
  to ++(1,0)
  to [european resistor, l=$Z_{m1}$,-*,fill=violet] ++ (2,0) coordinate(Port1)
  (Sec) to [short,-*] ++ (3,0)
   to ++(0,-0.5) coordinate(z_l)
  (Pri) ++(-0.25,0.5)
  node[anchor=west] {$1:N_1B_{11}$};
  
  \draw (3,-0.5) node[ocirc,scale=1.5, fill=black]{};
  \draw (3,-1) node[ocirc,scale=1.5,fill=black]{};
  \draw (3,-1.5) node[ocirc,scale=1.5,fill=black]{};
\draw
  (0,-4) to [open, v^>=$V_{in}$] (0,-2) 
  to [short, *- ,i=$I$] ++(2,0) coordinate(Cap_n)
  to [C, l_=$C_{01}$] ++(0,-2.5)
  to [short,-*] (0,-4.5) 
 (Cap_n) to ++(2,0) coordinate(Pri_n)
  to [L] ++(0,-2.5)
   to ++(-2,0);
  \draw
 (Pri_n) ++(1,-2.5) coordinate(Sec_n)
   to [L] ++(0,2.5)
  to ++(1,0)
  to [european resistor, l=$Z_{mN}$, -*,fill=violet] ++ (2,0)coordinate(Port1_n)
   (Sec_n) to [short,-*] ++ (3,0)
   to ++(0,-0.5) coordinate(z_l_n)
   (Pri_n) ++(-0.25,0.5)
  node[anchor=west] {$1:N_NB_{N}$}
  ;
\node[draw,minimum width=2cm,minimum height=5.7cm,anchor=south west, fill=cyan!25] at (z_l_n){ $[z_i]$}
(z_l_n) ++(0.25,0.5)
 node[anchor=west] {$b$}
   (Port1_n) ++ (0.25,0)
  node[anchor=west] {$a$}
  (z_l) ++(0.25,0.5)
 node[anchor=west] {$b$}
   (Port1) ++ (0.25,0)
  node[anchor=west] {$a$};
  \end{circuitikz}
  \end{center}
\caption{Multicell system equivalent circuit of N cells.}
\label{fig:EquivalentCircuit_array}
\end{figure}\

\begin{equation}
  z_i = z_{ii} + \sum_{j\neq i}^ N \frac{u_i}{u_j} z_{ij}
  \label{eq: acoustic_impadance_array}
\end{equation}\

\noindent where the first term represents the self impedance, and the second one is the mutual impedance. This shows the coupling effect that the neighbor cells produces on their velocities.

\section{DESIGN PROCESS \label{sec:Design_process}}
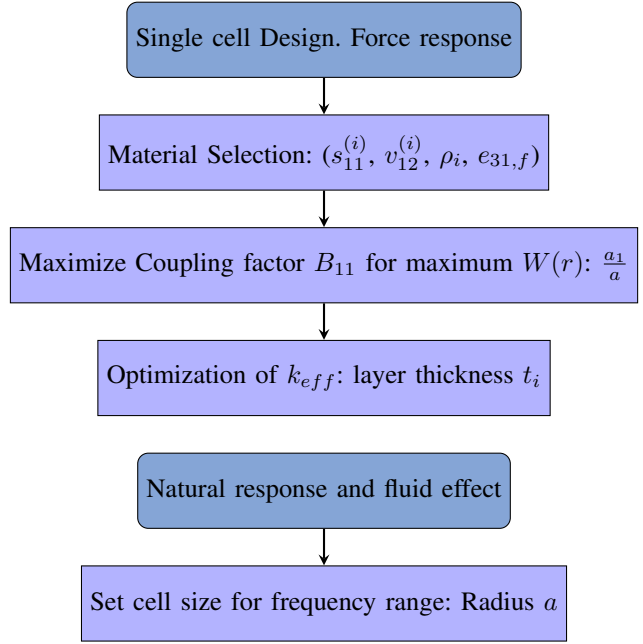
\begin{figure}[ht]
\begin{tikzpicture}[node distance=1.5cm]
\node(start)[startstop] {Single cell Design. Force response};
\node(pro1material)[process, below of = start] {Material Selection: ($s_{11}^{(i)}$, $v_{12}^{(i)}$, $\rho_i$, $e_{31,f}$)};
\node(pro2B11)[process, below of = pro1material] {Maximize Coupling factor $B_{11}$ for maximum $W(r)$: $\frac{a_1}{a}$};
\node(pro3keff)[process, below of=pro2B11]{Optimization of $k_{eff}$: layer thickness $t_i$};
\node(start2)[startstop, below of= pro3keff] {Natural response and fluid effect};
\node(pro4radius)[process, below of = start2] {Set cell size for frequency range: Radius $a$ };

\draw[arrow] (start) -- (pro1material);
\draw[arrow] (pro1material) -- (pro2B11);
\draw[arrow] (pro2B11) -- (pro3keff);
\draw[arrow] (start2) -- (pro4radius);
\end{tikzpicture}
    \caption{Flow chart for PMUT design. The steps are considered for a single cell only and divided in two milestones that separate the analysis of the force and frequency responses of the plate. Each box includes the design parameters that are critical for each step.}
    \label{fig: Flow_diagram}
\end{figure}\

The key steps involved in the design map of a single cell are represented in the flow chart of Fig. \ref{fig: Flow_diagram}. The design parameters have been grouped into two main categories. The first one is related to the force response, and the second one to the natural frequency response. For the first block, it is crucial to start by defining the materials by considering its elastic properties such as relevant compliance $s_{11}^{(i)}$ and the Poisson's ratio $v_{12}^{(i)}$, and in the case of piezoelectric materials, its characteristic constant $e_{31,f}$. Then, the maximum of the coupling factor $B_{11}$ is found with (\ref{eq:CouplingFactor}). Thus, the maximum displacement is reached when $B_{11}$ is equal to 1. This occurs for a ratio $a_{ac}= 0.67$. This stage includes an optimization block where the thickness of each layer determines the effective coupling coefficient. The most general form to calculate the effective electro-mechanical coupling coefficient $k_{eff}$ is defined in (\ref{eq:keff}):

\begin{equation}
    k_{eff}^2 = \frac{N_1^2 C_{m,1}}{N_1^2 C_{m,1} + C_{01}}
    \label{eq:keff}
\end{equation}\

Further manipulation yield to an expression that does not depend on the radius. It is assumed that $B_{11}$ was maximized in the previous step of the design, and thus $a_{ac} = \dfrac{a_1}{a}$ is known. The effective coupling coefficient in (\ref{eq:keff}) is now expressed in (\ref{eq: keff_new}) as:

\begin{equation}
    k_{eff}^2 = \frac{N_1^2 t_p}{N_1^2 t_p + 197.52 \pi^2 D \epsilon_{33}^\sigma a_{ac}^2 (1-k_{31}^2) }
    \label{eq: keff_new}
\end{equation}\

Following the flow chart, the effective coupling coefficient is computed with the materials described in Table \ref{tab:material_properties}. The contour plot of Fig. \ref{fig: effective_coupling} shows the values of the effective coupling coefficient for different thicknesses of the silicon and piezoelectric layers. To optimize $k_{eff}$, thickness for Si and PZT-4 layers can be selected in the last step in the force response block.

\begingroup
\setlength{\tabcolsep}{10pt} 
\renewcommand{\arraystretch}{1.5} 
\begin{table}[h!]
    \centering
    \caption{Material properties for $k_{eff}$ calculations \cite{muralt2005piezoelectric}.}
    \begin{tabular}{c c c c c}
    \hline
        t & Material & $s_{11}(m^2/N)$ & $\upsilon = \upsilon_{12}$ & $\rho_t (kg/m^3)$ \\ [0.75em]
        \hline 
        1 & Si  & 7.67 $\times 10^{-11}$ & 0.278 & 2330 \\
        2 & Si0$_2$ & 14.3$\times 10^{-11}$ & 0.3 & 2220 \\
        3 & Pt  &  7.34 $\times 10^{-11}$ &  0.420 & 21000 \\
        4 & PZT-4 & 13.2$\times 10^{-11}$ & 0.288 & 7700\\
        \hline
    \end{tabular}
    \label{tab:material_properties}
\end{table}\

\begin{center}
\begin{figure}[h!]
\centering
        \includegraphics[width=0.45\textwidth]{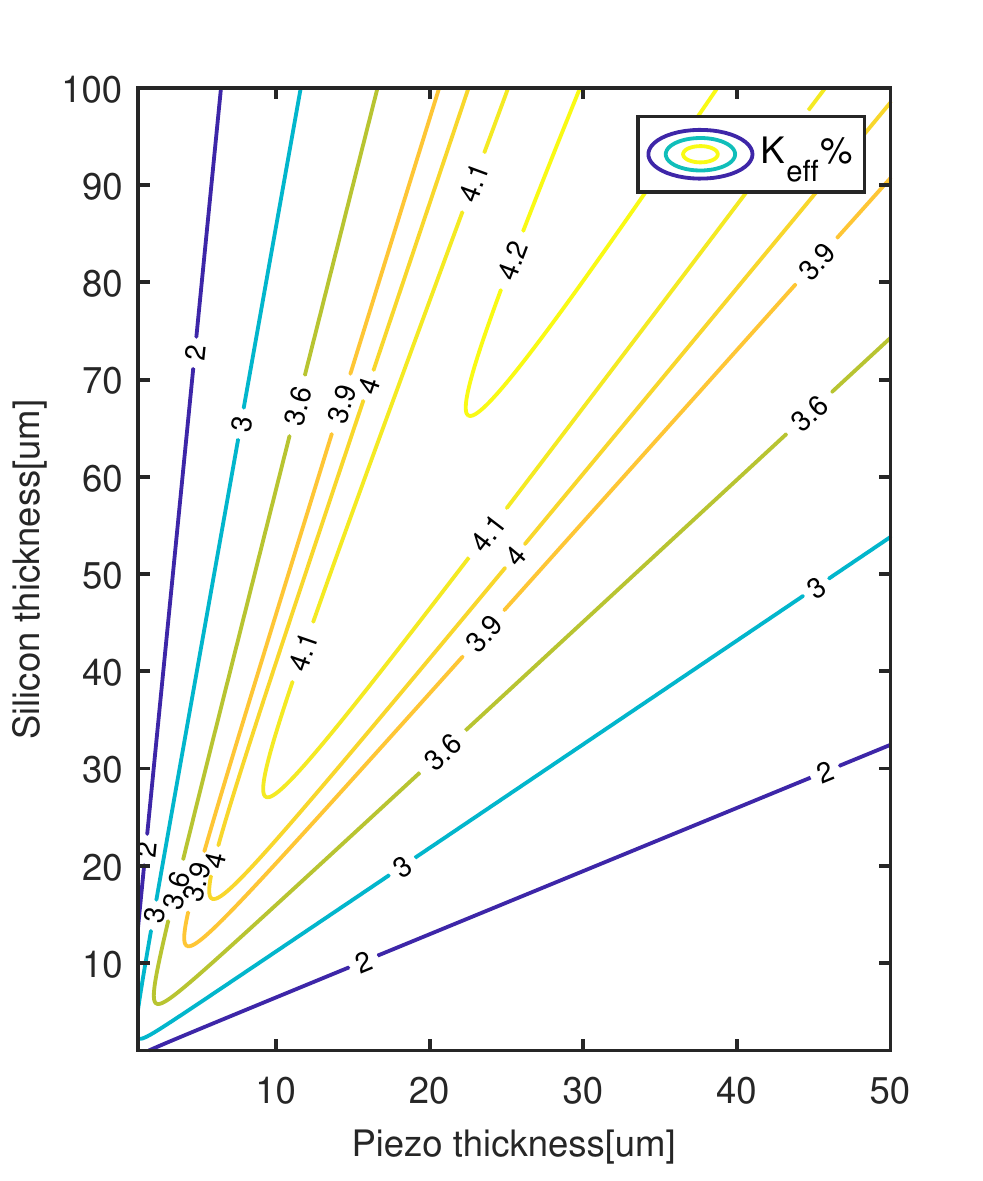}
    \caption{Effective coupling coefficient given by (\ref{eq:keff}). $Pt$ electrodes of 0.175 $\mu$m and $SiO_2$ thickness of 0.2$\mu$m are considered.}
    \label{fig: effective_coupling}
  \end{figure}
\end{center}\


\begin{figure*}[h!]
 \begin{subfigure}[t]{0.5\textwidth}
  \centering 
        \includegraphics[width=1\textwidth]{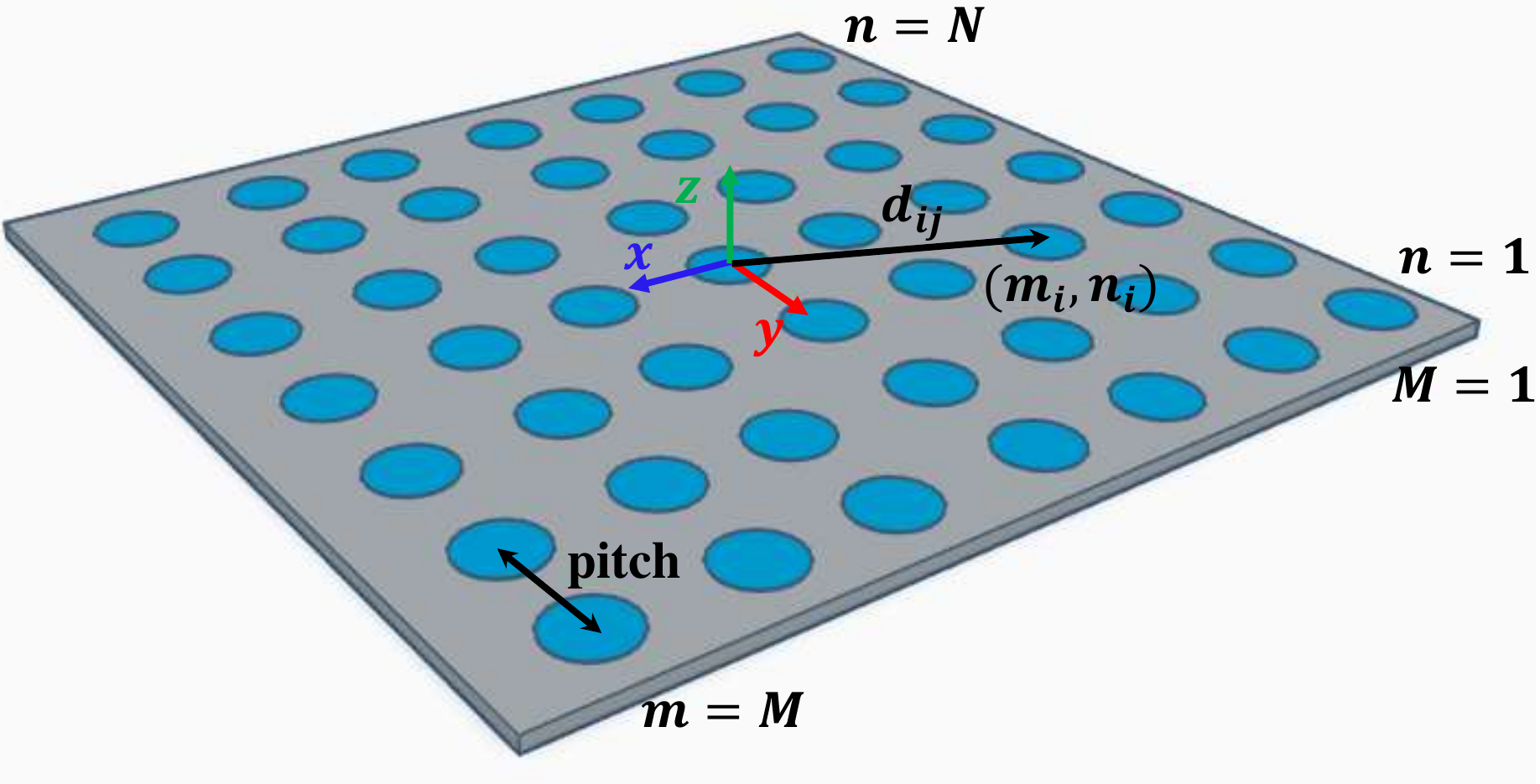}
         \caption{ }
         \label{subfig:geom_par}
     \end{subfigure}
\begin{subfigure}[t]{0.5\textwidth}
  \centering 
        \includegraphics[width=0.8\textwidth]{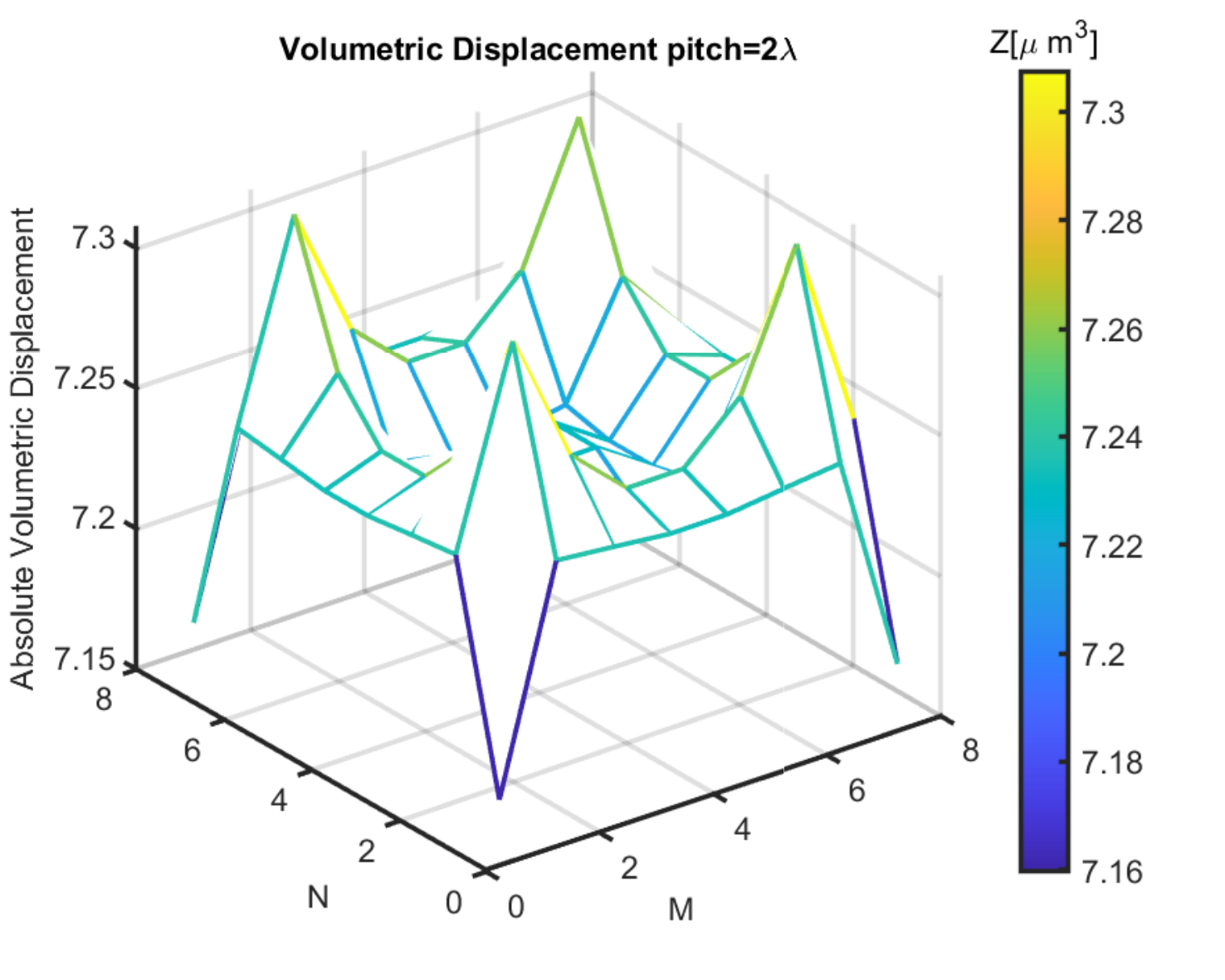}
         \caption{}
         \label{subfig:pitch_2lambda}
     \end{subfigure}
      \begin{subfigure}[t]{0.5\textwidth}
  \centering 
        \includegraphics[width=0.8\textwidth]{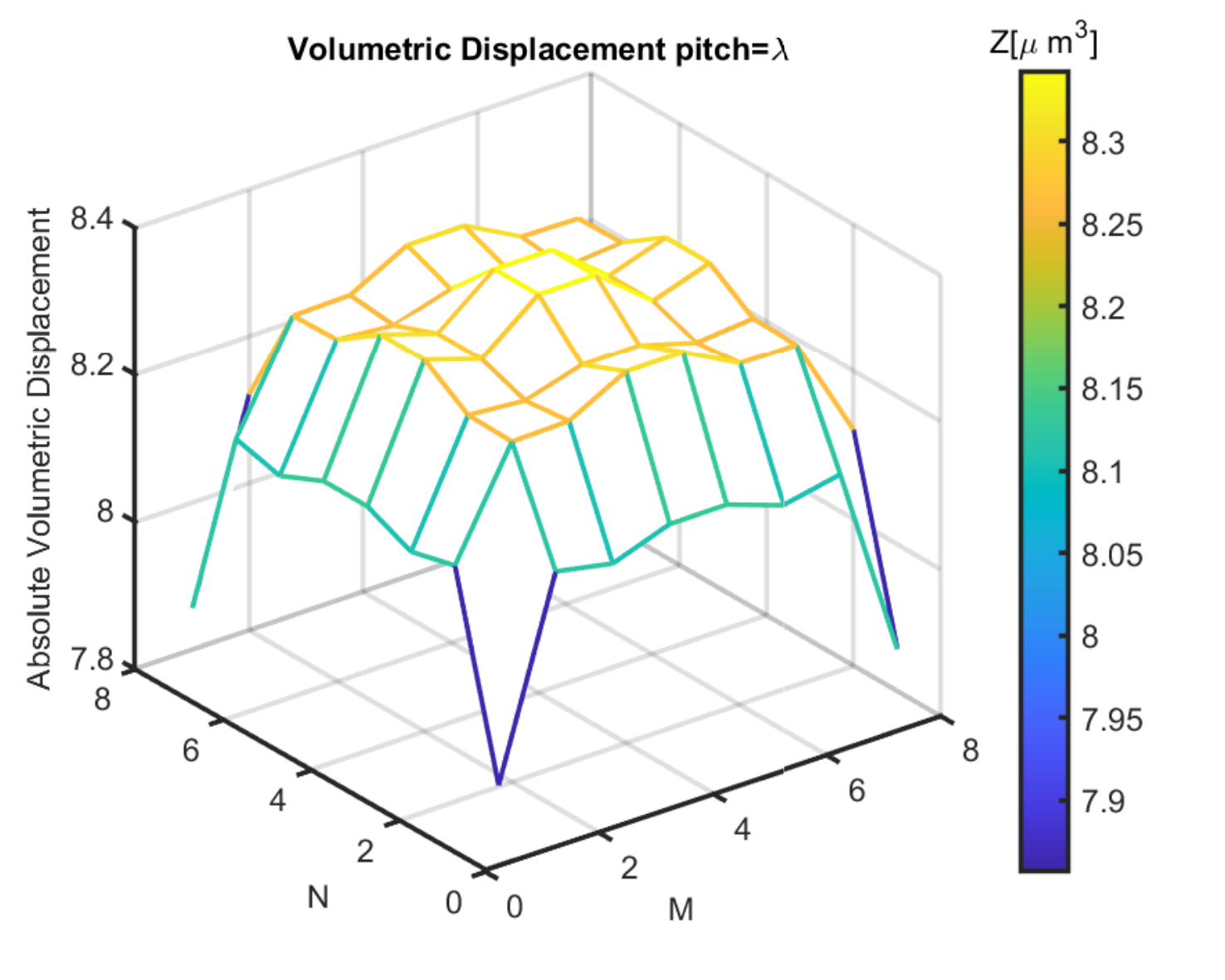}
         \caption{}
         \label{subfig:pitch_lambda}
     \end{subfigure}
          \begin{subfigure}[t]{0.5\textwidth}
  \centering 
        \includegraphics[width=0.8\textwidth]{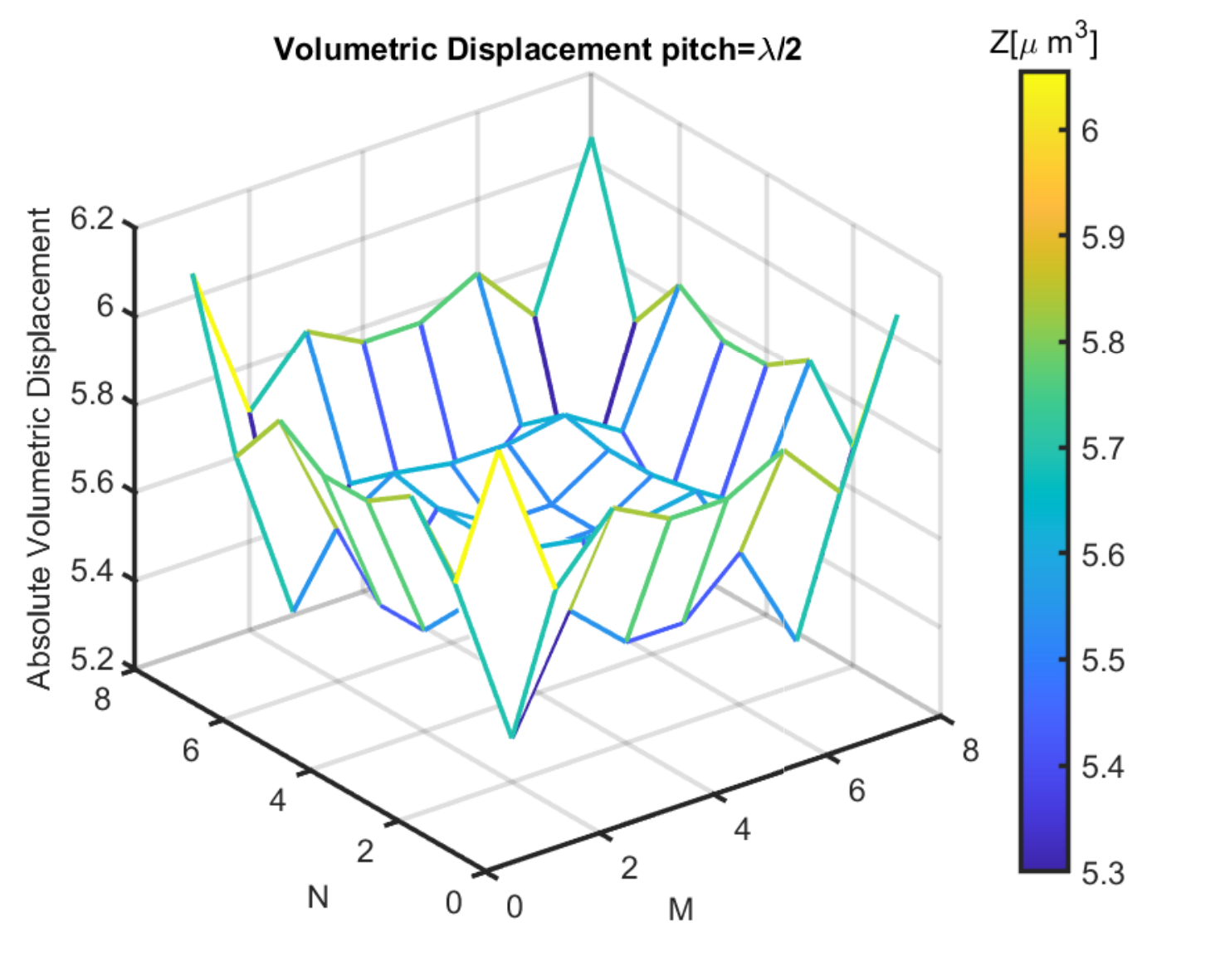}
         \caption{}
         \label{subfig:pitch_lambdaHalf}
     \end{subfigure}
  \caption{Analysis of the distance(pitch) between adjacent cells. (a) Schematic with the geometrical parameters of a PMUT array. Volumetric displacement for an array of 8 x 8 number of cells with (b) pitch $=2 \lambda$, (c) pitch $= \lambda$, and (d) pitch $= \frac{\lambda}{2}$.}
  \label{fig: estimate_pitch}
\end{figure*}\
The second block of the design flow process includes the natural frequency response where the resonance frequency is found. Implying that the uni-morph structure behaves like a plate, the radius of the plate is found with (\ref{eq:frequency}).\\

\subsection{Wireless Power Transfer scheme\label{subsec:WPT}}
In this section, we aim to design the wireless power transfer system that will be used to power a brain implant as it was described in the introduction. First, a highly effective coupling coefficient is ensured by the correct selection of the layers of silicon, and PZT-4. Then, the radius of the cell is calculated with expression (\ref{eq:frequency}). The structural parameters of the multilayered plate to be used for a wireless power transfer system is summarized in Table \ref{tab:structural_dimensions} and has been shown in Fig. \ref{fig: PMUT_schematic}. The pressure values get from a PMUT cell is low due to its small dimension. This forces a connection in parallel of cells to form a larger array that achieves the practical values of acoustic intensity which is crucial for wireless power transfer applications.\\

\begingroup
\setlength{\tabcolsep}{10pt} 
\renewcommand{\arraystretch}{1.5} 
\begin{table}[h!]
    \centering
    \caption{Structural parameters of the analyzed PMUT.}
    \begin{tabular}{c c c c}
    \hline
        Symbol & Value & Unit & Description \\ [0.75em]
        \hline 
        $t_1$ & 10   & $\mu$m & Silicon thickness\\
        $t_2$ & 0.2 & $\mu$m & Silicon Dioxide thickness\\
        $t_3$ & 0.175  & $\mu$m & Bottom electrode thickness\\
        $t_4$ & 5  & $\mu$m & Piezoelectric thickness \\
        $t_5$ & 0.175  & $\mu$m & Top electrode thickness \\
        $a$   & 107 & $\mu$m & Diaphragm radius  \\
        $a_1$ & 71 & $\mu$m & Top electrode radius\\
        $k_{eff}$ & 4 $\%$ & & Effective coupling coefficient\\
        \hline
    \end{tabular}
    \label{tab:structural_dimensions}
\end{table}
To start with the array design, the Rayleigh length $\dfrac{A}{\lambda}$ is used to estimate the area ($A$) of the transducer. The area will define the location of the maximum acoustic intensity. Therefore, the number of cells and the distance between adjacent PMUTs are the parameters to be established for the array design. An analysis can be done by using an equivalent circuit model and a mutual impedance calculation as it is expressed in \cite{pritchard1960mutual}. This work presents a method to obtain the mutual impedance through infinite series as it is depicted from equation (\ref{eq:mutualImpedancenm}) to (\ref{eq:zetaNM}): 
\begin{align}
Z_{12} &= \rho c \pi a^2 \sum_{p=0}^\infty \sigma_p(ka)\cdot \left( \frac{a}{d_{ij}} \right)^{p} \cdot \zeta_p (kd_{ij})\label{eq:mutualImpedancenm}\\[0.75em]
\sigma_p(ka) &= \frac{2\Gamma(p+\frac{1}{2})}{\pi^{\frac{1}{2}}} \sum_{n=0}^p \frac{J_{n+1}(ka) J_{p-n+1}(ka)}{n!(p-n)!}\\[0.75em]
\zeta_{p} &= \left( \frac{\pi}{2kd_{ij}}\right) [ J_{p +\frac{1}{2}}(kd_{ij}) + j(-1)^{p} J_{-m-n-\frac{1}{2}}(kd_{ij})]\label{eq:zetaNM}
\end{align}\

\noindent where $p = m+n$, $\Gamma$ and $J_{p+\frac{1}{2}}$ are the gamma and complex spherical Bessel functions, respectively. The model is established under the assumption that all circular plates in the array are vibrating in phase. This numerical model and the volumetric velocity matrix equation of \cite{akhbari2016curved} are used to estimate the pitch ($d_{ij}$) by computing the volumetric displacement. Fig. \ref{fig: estimate_pitch} presents the geometrical parameters of a square PMUT array and an analysis of the volumetric displacement for three different pitches. The highest displacement is obtained for a pitch equals $\lambda$, which is shown in Fig.\ref{fig: estimate_pitch} (\subref{subfig:pitch_lambda}). This is explained by the fact that all cells vibrate at the same phase, and for an odd integer number of the wavelength this condition is reached. When the distance between two cells is not an odd integer number, the phase shift between waves yields constructive or destructive interference bringing about changes in the velocity of their neighbor cells. The minimum displacement is given for a pitch equal to $\dfrac{\lambda}{2}$.\\
The second calculation is done for the acoustic intensity by taking the approach of \cite{smyth2017piezoelectric}. Therefore, the model of frequency-dependent attenuation of (\ref{eq: attenuation}) detailed in \cite{wygant2008integration},\cite{wygant2011comparison} will be taken into account in the design:\\

\begin{equation}
    P = P_0 e^{-\alpha f^\kappa Z}
    \label{eq: attenuation}
\end{equation}\

\noindent where $\alpha$ and $\kappa$ are constants of attenuation given for the propagation medium. $P_0$ is the pressure on the transducer surface, $f$ is the resonance frequency and $Z$ the propagation distance. Assuming that the attenuation can be neglected at a maximum distance of 4 mm, the pressure is defined by considering just diffraction losses as it is reported in \cite{wygant2008integration}, \cite{wygant2011comparison}. Equation (\ref{eq:prressure_z}) represents the analytical model of this approximation:\\

\begin{equation}
    P(Z) = \left [ \frac{\rho c |u|^2 R_r}{2\pi}\right]^{\frac{1}{2}} \frac{1}{Z}
    \label{eq:prressure_z}
\end{equation}\
\noindent where $u$ is the average velocity of the diaphragm and $R_r$ is the real part of the radiation impedance given in (\ref{eq: radiation_impedance}).\\

Then, the actuation efficiency is found by having the transfer function, when the transducer is operating at its fundamental resonance frequency. By replacing the average velocity, the transfer function is represented in (\ref{eq: sensitivity}):\\

\begin{equation}
    G_t = \frac{P(Z)}{V_{in}} = \frac{1}{2\pi k a^2 } \frac{N_1}{Z}
    \label{eq: sensitivity}
\end{equation}\

As it is well established in \cite{jung2017review}, the actuation efficiency is directly proportional to  $e_{31,f}$ as a material-dependent parameter. In this context, a piezoelectric layer PZT-4 was selected for the wireless power transfer scheme. In terms of the geometry, the transmitter actuation efficiency is inversely proportional to the square of the radius and directly proportional to $Z_p$. 
In order to obtain an intensity close to the maximum and the target distance, a multi-cell model is used. A brief estimation of the number of cells can be calculated with equation (\ref{eq:number_cells}):\\

\begin{equation}
    K = \frac{2\pi P^2(Z) Z^2 R_r}{\rho_0 c_0 (N_1V_{in})^2}
    \label{eq:number_cells}
\end{equation}\

\noindent where resonance in the fundamental mode is assumed, and all cells contribute equally to the output power. This equation is used to plot the values of $G_t$ for the pitches of maximum and minimum volumetric displacements ($\lambda$ and $\frac{\lambda}{2}$). Although a higher displacement amplitude can be reached for a pitch equal to $\lambda$, the number of cells is limited by the constraint in the area through the Rayleigh distance of the application. By plotting the actuation efficiency in Fig. \ref{fig:Pressure_vs_M_z}, a higher number of cells delivers more pressure per voltage for a transducer of the same area. Therefore, the distance between adjacent cells was reduced to a compact configuration with a pitch equals $\frac{\lambda}{2}$. For a transducer area of 3.24 mm$^2$, a total of 49 cells is simulated. A Rayleigh distance of 6 mm is selected, longer than the requirement. This is justified by the fact that the maximum intensity occurs in the near field, and the Rayleigh model does not predict accurately the focal points for arrays due to its discrete contribution. However, the model presents a significant deviation for arrays due to the discrete contribution of each cell as it is mentioned in \cite{akhbari2015bimorph}.

\begin{figure}[t!]
    \centering
    \includegraphics[width=0.5\textwidth]{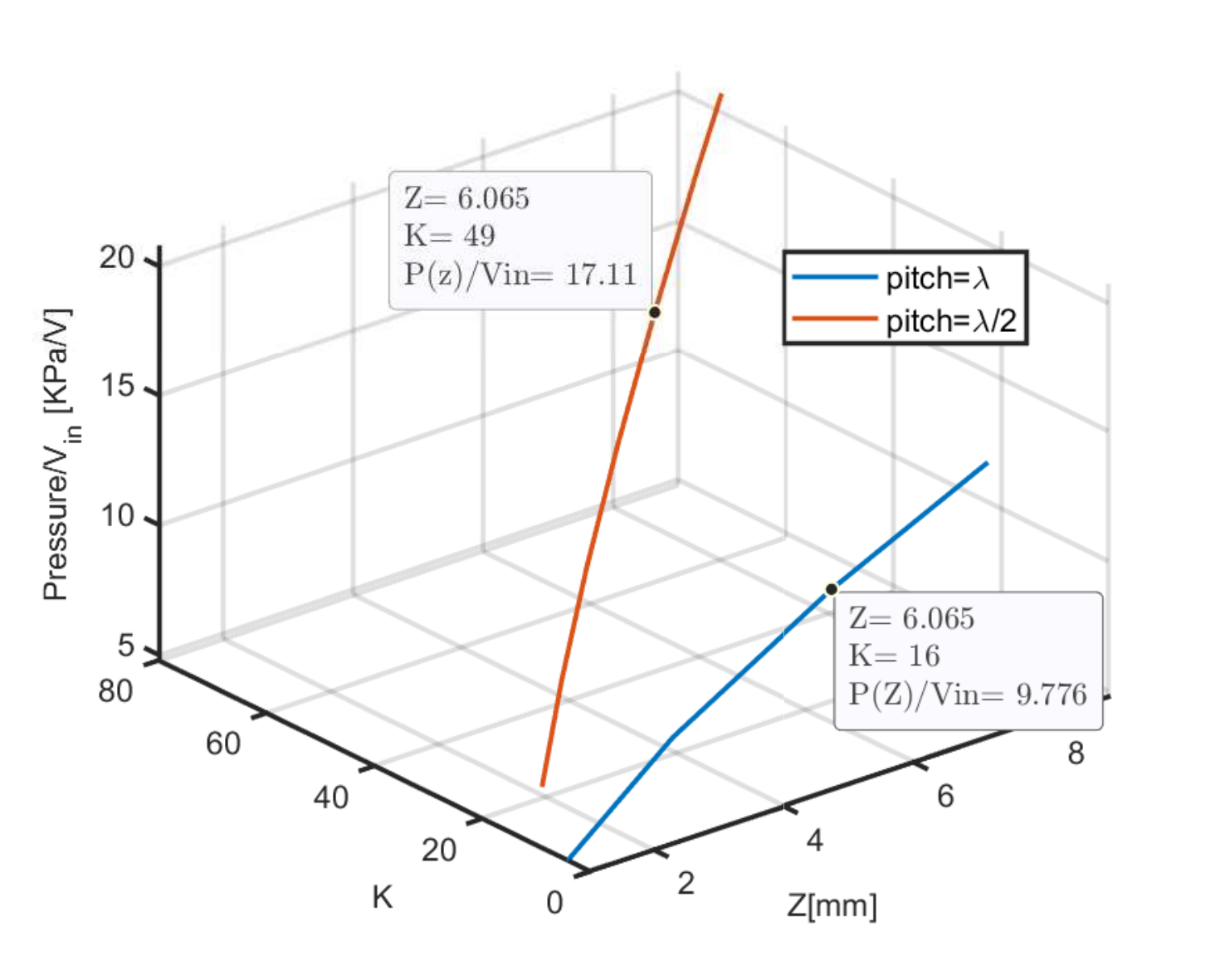}
    \caption{$G_t$ vs number of cells $K$ and distance $Z$. The points where the area of the transducer is 3.24mm$^2$ have been marked.}
    \label{fig:Pressure_vs_M_z}
\end{figure}
The parameters of the PMUT array to be analyzed are summarized in Table \ref{tab:array_dimensions}. The number of cells was selected to yield approximately an active area of the transducer that has a Rayleigh distance close to 6 mm and a distance between adjacent PMUTs equal to half of the wavelength.\\

\begingroup
\setlength{\tabcolsep}{10pt} 
\renewcommand{\arraystretch}{1.5} 
\begin{table}[h!]
    \centering
    \caption{Summary of the structural parameters of the analyzed PMUT array.}
    \begin{tabular}{c c c c}
    \hline
        Symbol & Value & Unit & Description \\ [0.75em]
        \hline 
        $d_{ij}$ & 260  & $\mu$m & Distance between cells\\
        $K$ & 49 &  & Total number of cells\\
        $M$ & 7  &  & Number of cells in a row\\
        $N$ & 7  &  & Number of cells in a column \\
        $A$ & 3.24  & mm$^2$ & Array area \\
        $Z$ & 6.065  & mm & Rayleigh distance \\
        \hline
    \end{tabular}
    \label{tab:array_dimensions}
\end{table}

\section{SIMULATION RESULTS\label{sec:Simutation_res}}
In this section, COMSOL Multiphysics V5.4 is used as the finite element method (FEM) simulation tool to verify the compact theoretical design. The geometrical parameters of the array are presented in Table \ref{tab:structural_dimensions} and \ref{tab:array_dimensions}. The material properties are illustrated in Table \ref{tab:material_properties}. To find the acoustic profile, PMUT is 3-D modeled in water domain . Physics of solid mechanics, electrostatics, and electric currents are employed for solving equations in electro-mechanical interface of the PMUT, and pressure acoustics in frequency domain is used for computing the wave equation in water domain. 

Regarding the boundary conditions, fixed constraint on the edge of each PMUT cell in solid mechanics is utilized. In electrostatics, Pt bottom layer and Pt top layer are considered as bottom and top electrodes, respectively. To simplify the model and lowering the cost of computation, 3-D profile of top electrodes of each cell are considered as 2-D shell layer in electric currents physic. Finally, for pressure acoustics, the spherical wave radiation is selected on the outer domain of water for absorbing the reflected waves. To couple these physics, a multiphysics node is added with subnodes of piezoelectric effect for interfacing the solid mechanics and electrostatics, and acoustic-structure boundary for interfacing the solid mechanics with pressure acoustics.
\begin{figure}[h!]
    \centering
    \includegraphics[width=0.5\textwidth]{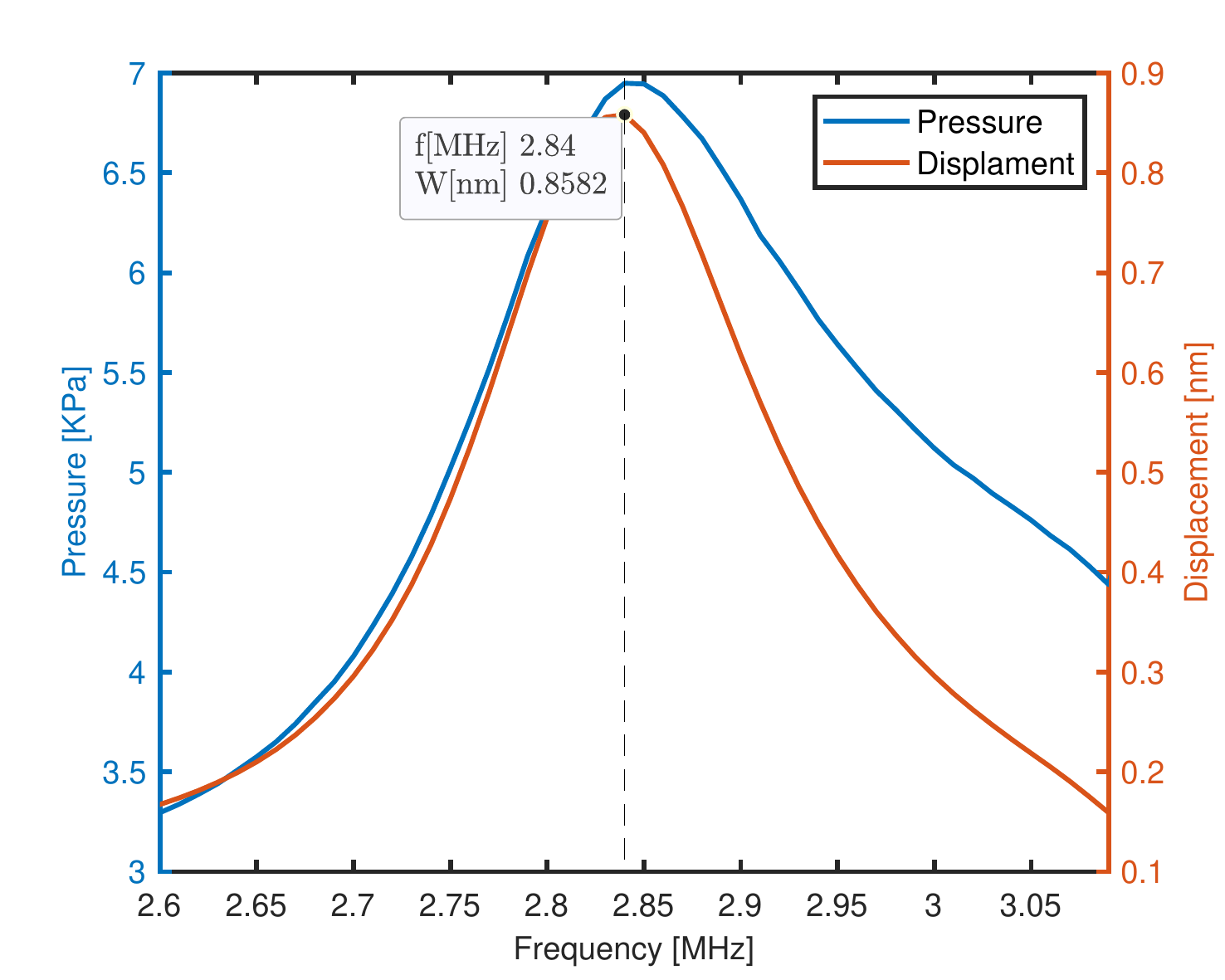}
    \caption{Simulation of pressure and displacement in the center of the PMUT vs frequency.}
    \label{fig:Pressure_vs_frequency}
\end{figure}\

\subsection{Single cell}
Initially, the resonance frequency of the design is simulated. The values of the displacement and pressure are measured in the center of the PMUT cell for a parametric sweep of the frequency and a driving voltage equals to 1 V. Fig. \ref{fig:Pressure_vs_frequency} represents the frequency response of the PMUT cell. The design results in a displacement and pressure that have a peak at 2.84 MHz around 5 $\%$ of deviation from the required value of 2.7 MHz. As a result, the displacement is 0.8582 nm. The maximum pressure is obtained at the same frequency with a peak value of 6.89 KPa. 

 Further analysis was made to obtain the maximum voltage for safe tensile stress. For the case of PZT-4, the peak of tensile stress is 24 MPa\cite{sherman2007transducers}. Since the PMUT design consists of a thin layer, the analysis of  \cite{muralt2005piezoelectric} is assumed. This establishes that thin PZT layers support ten times larger fields than the bulk materials and hence, piezoelectric
stresses become ten times larger. With this in mind, COMSOL is used to estimate the safe operating voltage of the design. 
\begin{figure}[h!]
    \centering
    \includegraphics[width=0.5\textwidth]{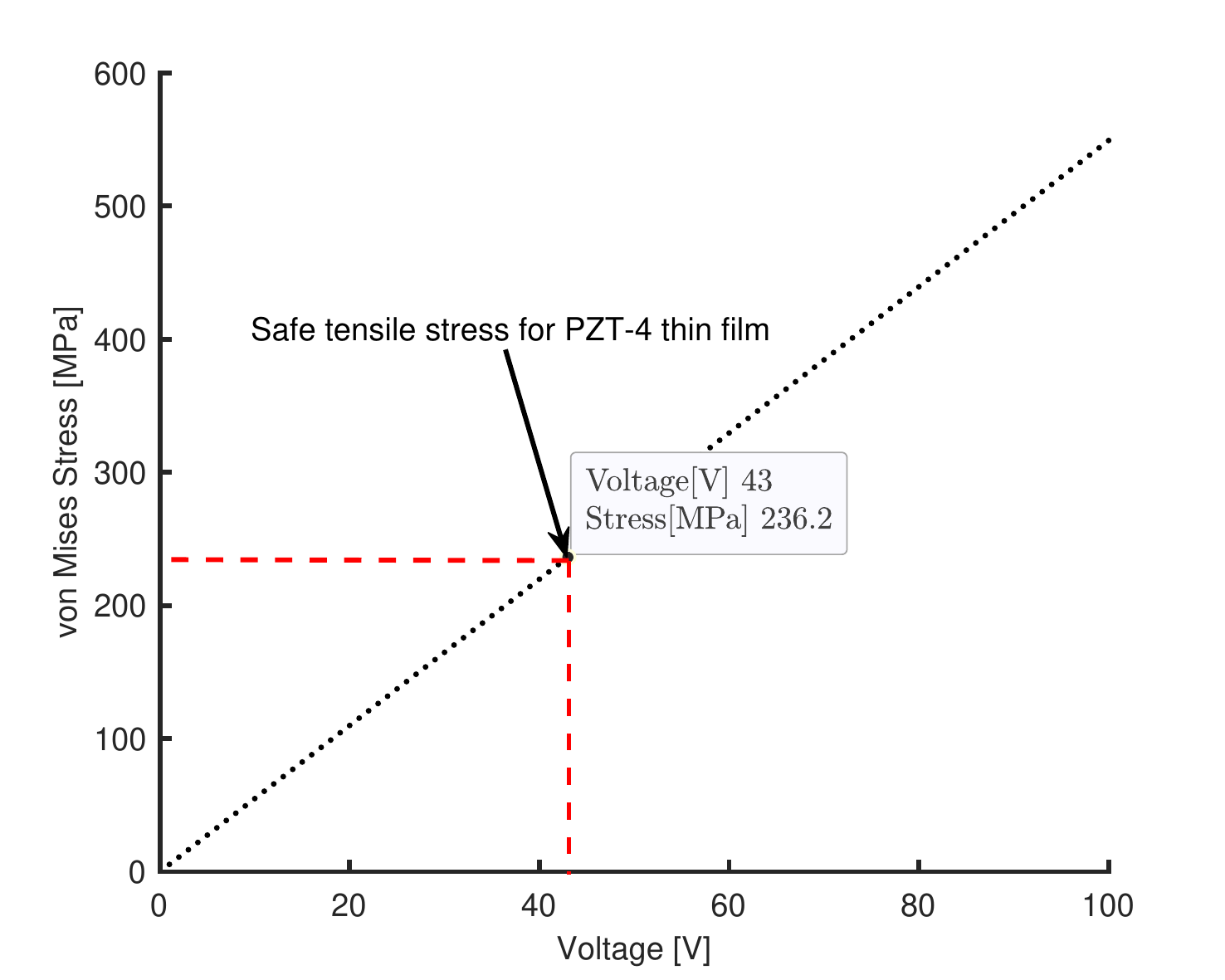}
    \caption{Simulation results of the von Misses Stress vs voltage amplitudes}
    \label{fig:stress_voltage}
\end{figure}

Figure \ref{fig:stress_voltage} shows that for a voltage of 43 V, the cell reaches a Von Misses Stress close to ten times the safe tensile stress of PZT-4. Therefore, this design can operate at a maximum voltage of 43 V before the buckle losing all load-bearing capability.

\subsection{Array}
Without beam steering, the maximum acoustic intensity occurs on-axis. Thus, plotting the intensity as a function of Z, a maximum is observed in the curves shown in  Fig.\ref{fig: Analysis_Int_z}(\subref{subfig:intensity_vs_z}) for different voltage amplitudes. Sound pressure levels of Fig. \ref{fig: Analysis_Int_z} (\subref{subfig:sound_levels}) show the acoustic distribution and a graphic estimation of the region of maximum acoustic intensity. 

\begin{center}
\begin{figure}[h!]
\centering
  \begin{subfigure}[t]{0.5\textwidth}
  \centering 
        \includegraphics[width=\textwidth]{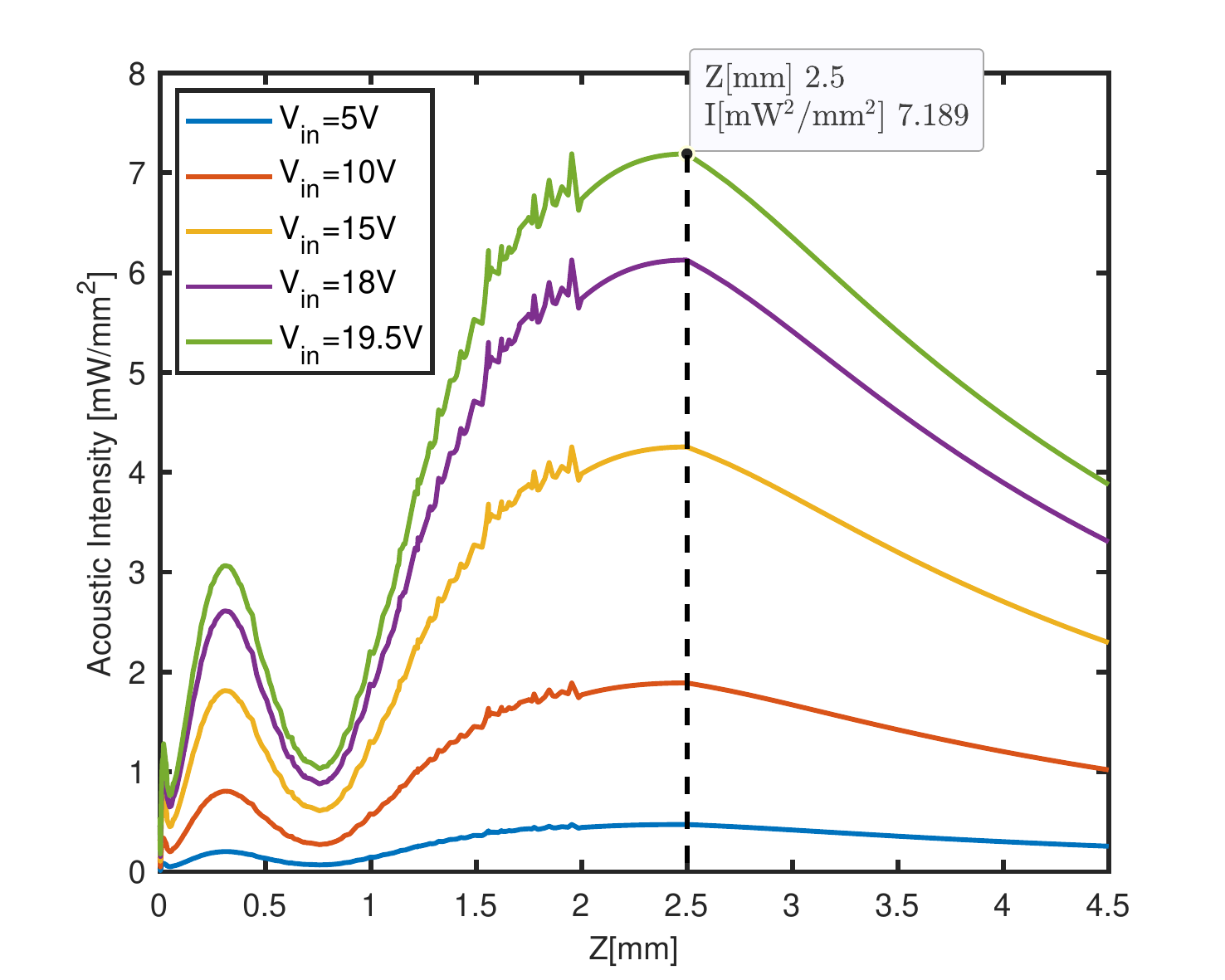}
        \caption{ } \label{subfig:intensity_vs_z}
    \end{subfigure}
    \begin{subfigure}[t]{0.5\textwidth}
        \centering 
        \includegraphics[width=\textwidth]{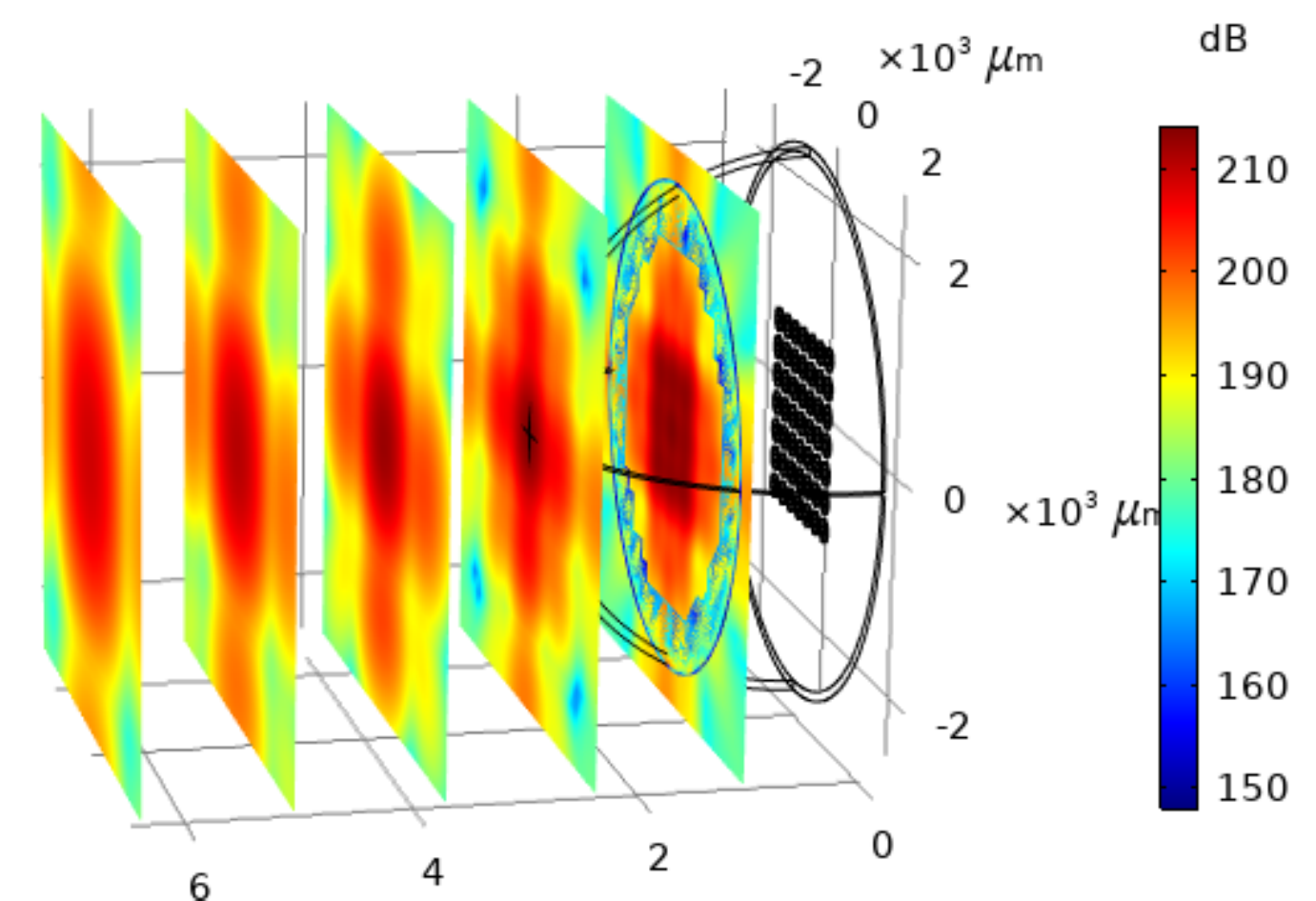}
        \caption{}\label{subfig:sound_levels}
    \end{subfigure}
    \caption{Simulated results for COMSOL Multiphysics simulation for (a) on-axis acoustic intensity, and (b) 3D simulation of sound pressure levels.}
    \label{fig: Analysis_Int_z}
  \end{figure}
\end{center}

The maximum acoustic intensity is marked at 2.5 mm, and its amplitude is changed by the applied voltage. To reach 7.189 mW/mm$^2$, which is the closest value to the FDA-allowed time-averaged acoustic intensity, a voltage of 19.5 V was applied as it is shown in the green curve of Fig.\ref{fig: Analysis_Int_z}(\subref{subfig:intensity_vs_z}). The result gives an upper voltage limit of 19.5 V to guarantee safe levels of intensity on the human tissue. Although duty-cycling the input signal, provide more flexibility for stronger acoustic intensity. 
\section{CONCLUSION}
This paper presents an analytical design of an ultrasound wireless power transfer system based on PMUT for powering sub mm-sized brain implants. With a goal of powering the brain implants, specific requirements have been considered such as the distance of maximum intensity, the value of the maximum intensity, and the resonance frequency in brain tissue. By following the compact flowchart presented in this paper, an optimization process of the effective coupling coefficient and frequency has been analyzed for a single cell transducer. Furthermore, the design of an array was carried out where the targeted distance and intensity were the main constraints in the process. An analysis of the mutual impedance was elaborated as a key step to optimize the array design. However, the constrain in the area (due to the target distance value) limits the design to a compact configuration with a pitch of half of the wavelength. The simulations show a close agreement between the analytical design and the results from COMSOL. The frequency is around 2.8 MHz and an array of 7 x 7 PMUTs is able to deliver 7.189 mW/mm$^2$, 2.5 mm away from the transmitter for 19.5 V. 

To conclude, a more reliable design can be put under test by using the approach following in this paper. Hence, design time of trial and error approaches decreases. 
\section{FUTURE WORK}
In the future, an analysis of a transceiver by assigning selectively one or multiple cells of the array as a receiver will be added. In addition, the experimental study of thin PZT layers and its safety tensile stress that limits the applied voltage will be investigated. Also, increasing the applied voltage by using a modulated signal requires further analysis. Lastly, a comparison between the square and circular arrays of PMUTs will be investigated. 
\section*{Acknowledgment}
This project (STARDUST) has received funding from the European Union’s Horizon 2020 research and innovation program under grant agreement No 767092.

\ifCLASSOPTIONcaptionsoff
  \newpage
\fi


\bibliographystyle{IEEEtran}
\bibliography{TBiocas_2019.bib}


\begin{IEEEbiography}[{\includegraphics[width=1in,height=1.25in,clip,keepaspectratio]{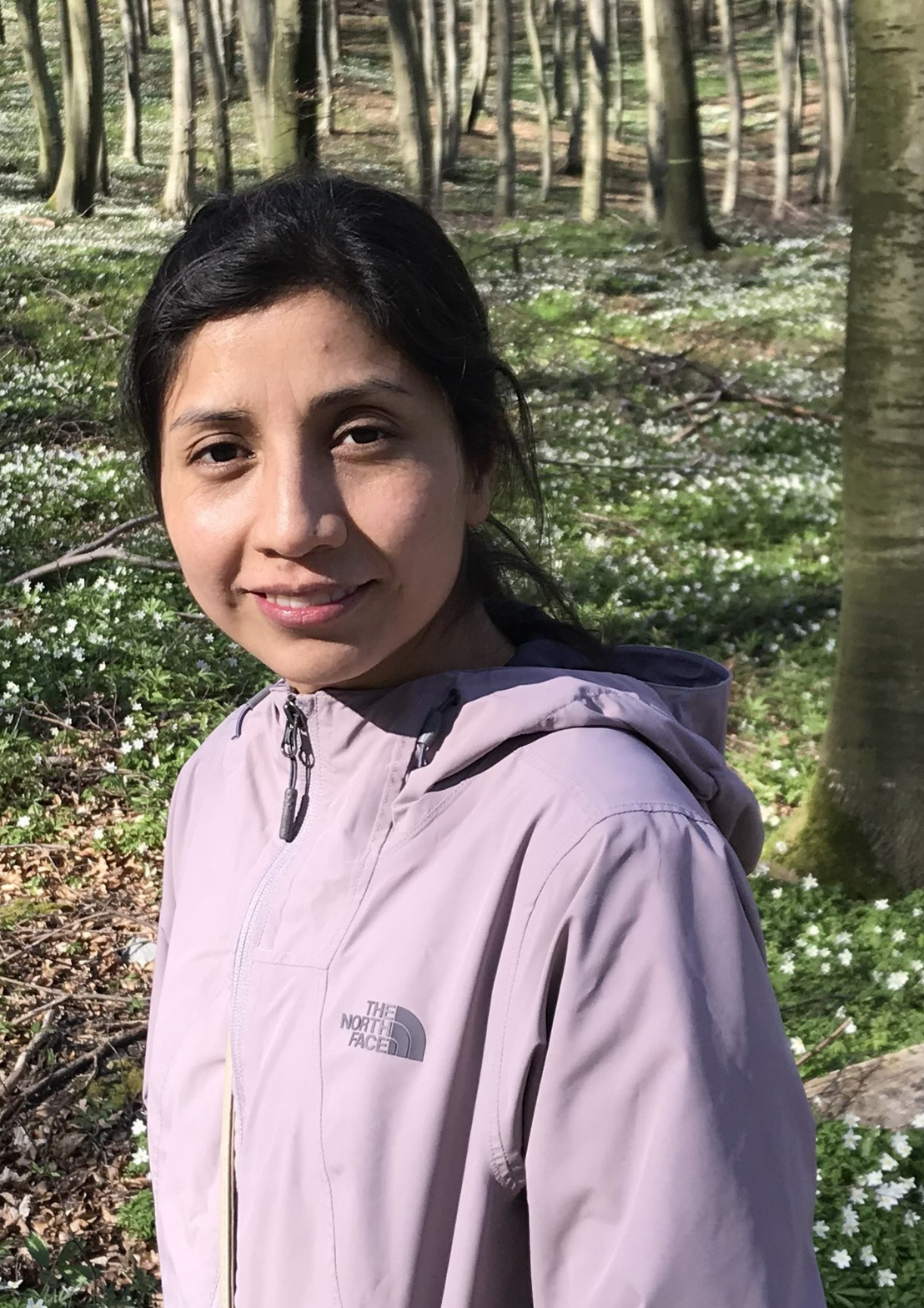}}]{Fernanda Narv\'aez}received the B.Sc. degree in electronics and control systems engineering from from the National Polytechnic School, Ecuador in 2010. In 2012 she received a scholarship from the Ecuadorian government, and a year after that she received a master’s degree in Sensors for Industrial Applications from the Polytechnic University of Valencia, Spain. She received a second Master's degree in Electrical engineering from Aarhus University, in 2020. Her current reseach interest include medical applications, microelectro-mechanical systems(MEMS). 
\end{IEEEbiography}

\begin{IEEEbiography}[{\includegraphics[width=1in,height=1.25in,clip, keepaspectratio]{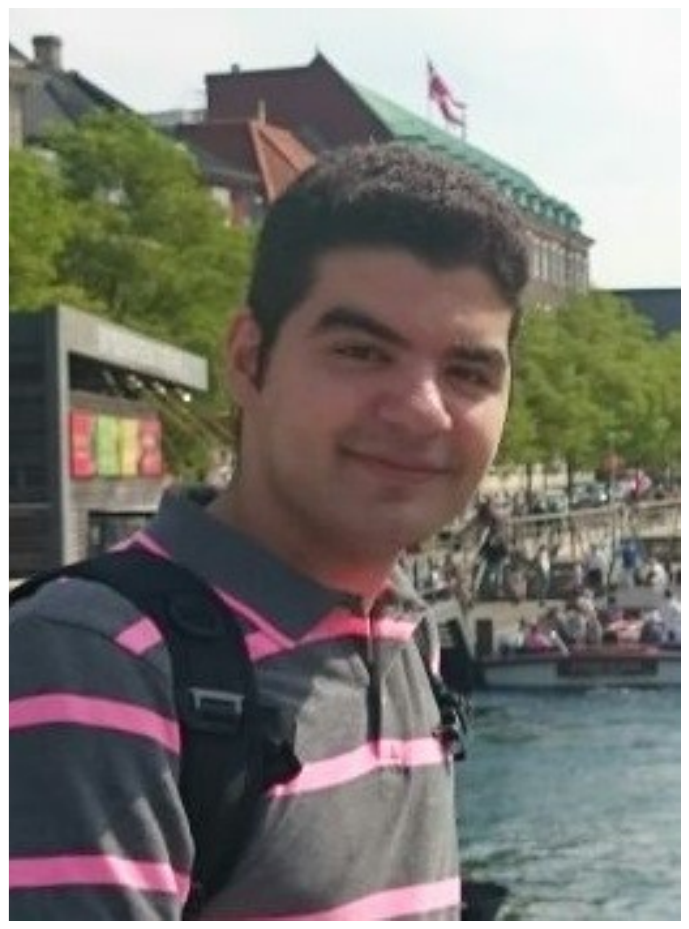}}]{Seyedsina Hosseini}received the B.Sc. degree in mechanical engineering from Azad University of Tehran, Iran in 2012. Due to his enthuthisasm in multidisciplinary fields, he pursued the master in Micro-and nanoelectromechanical Systems (MEMS\&NEMS) engineering at University of Tehran, Iran and obtained his degree in 2016. He is currently with ICE-Lab, Aarhus University, as a Ph.D. fellow where he is working on MEMS-based piezoelectric ultrasonic transducers for brain implants.
\end{IEEEbiography}

\begin{IEEEbiography}[{\includegraphics[width=1in,height=1.25in,clip,keepaspectratio]{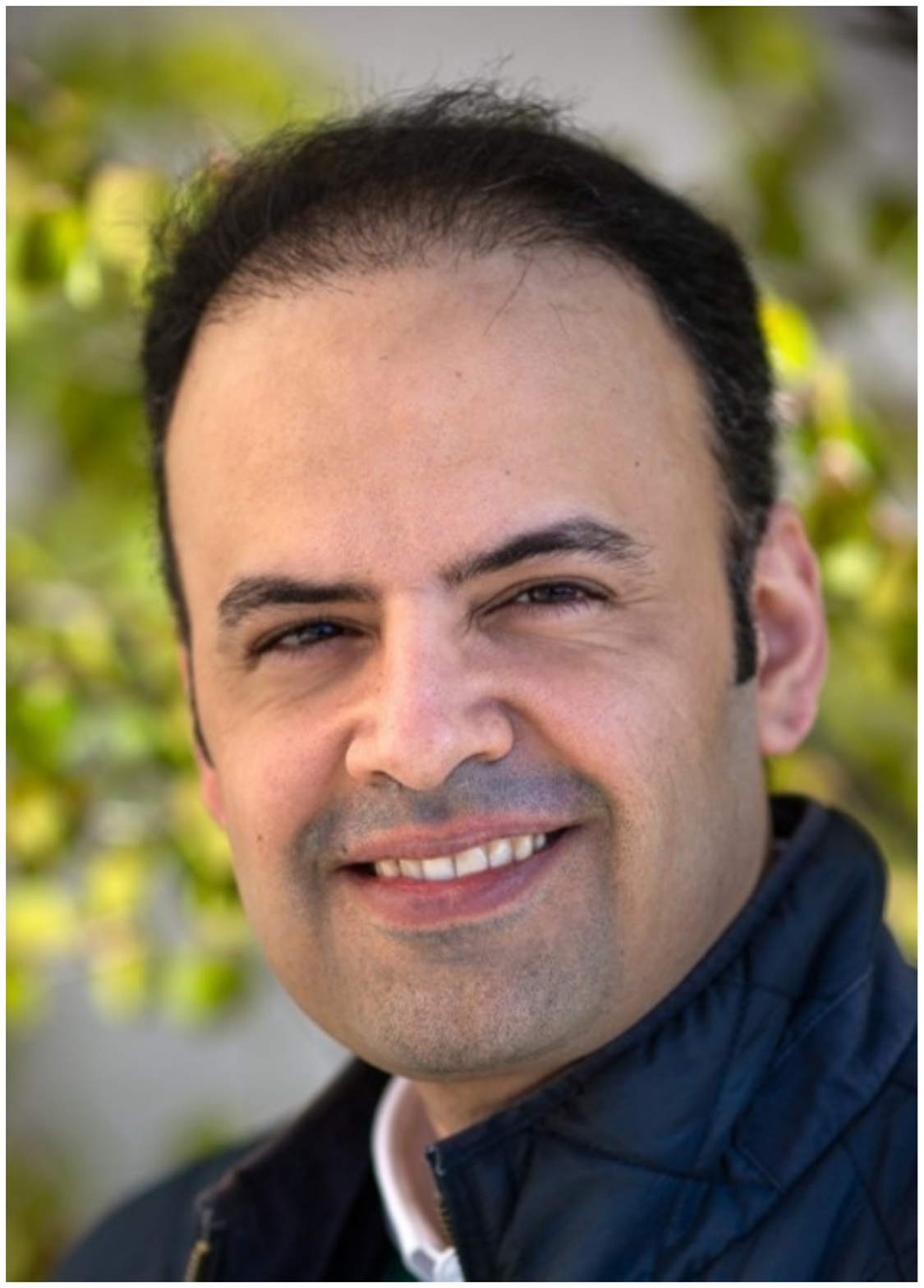}}]{Hooman Farkhani}
received the B.Sc. degree in electrical engineering from the University of Kashan,
Kashan, Iran, in 2004, and the M.Sc. and Ph.D. degrees in electronic engineering from the Ferdowsi University of Mashhad, Mashhad, Iran, in 2008 and 2014, respectively. He was a MSCA-IF post-doctoral at Aarhus University for two years from 2017 to 2019. He is currently an Assistant Professor with the Integrated Circuit and Electronics Laboratory, Department of Engineering, Aarhus University, Denmark. His current research interests include low power and low voltage spintronic-based Neuromorphic computing systems, STT-RAM design, SRAM design, and fully digital ADCs.
\end{IEEEbiography}

\begin{IEEEbiography}[{\includegraphics[width=1in,height=1.25in,clip,keepaspectratio]{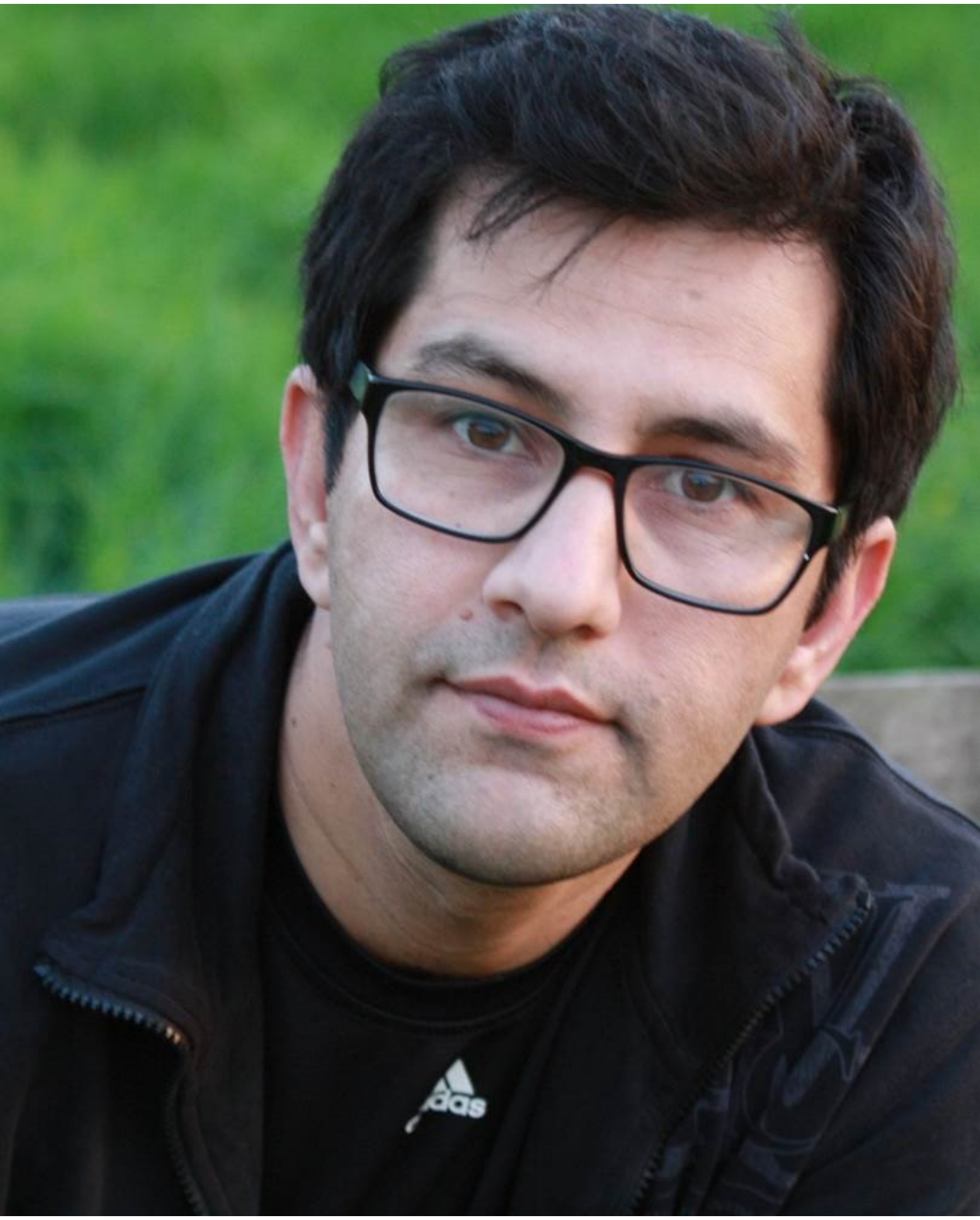}}]{Farshad Moradi (M'08, SM'18)}
received the BSc and MSc degrees electrical engineering from Isfahan University of Technology and Ferdowsi University of Mashhad, respectively. He received the PhD degree in electrical engineering
from University of Oslo, Norway, in 2011. From 2009 to 2010, he visited the Nanoelectronic Laboratory, Purdue University, USA. He started his academic career as an assistant professor at the
school of engineering at  Aarhus university and currently an ssooiate professor with the department of engineering at Aarhus University. He is is the director of the Integrated Circuit and Electronics Laboratory (ICELab). He is reviewer of many Journals and has been served as technical committee of several conferences. He is an associate editor of Integration, the VLSI and VLSI Journal. He is the author/co-author of more than 80 Journal and Conference papers. His current research interests include ultra low-power integrated electronics from device to architecture.
\end{IEEEbiography}

\vfill



\end{document}